\documentclass[aps,prd,twocolumn,showpacs,superscriptaddress,groupedaddress,nofootinbib]{revtex4}
\usepackage{xcolor}
\usepackage{graphicx,color,bm}

\usepackage{amsmath,amssymb}
\usepackage{epstopdf}
\usepackage{ulem}
\usepackage{array}
\usepackage{verbatim}
\usepackage[utf8]{inputenc}
\usepackage{rotating}
\newcolumntype{K}[1]{>{\centering\arraybackslash}m{#1}}

\usepackage[colorlinks,linkcolor=blue,citecolor=blue]{hyperref}

\begin{document}

\title{Cosmic expansion parametrization: Implication for curvature and $\text{H}_{0}$ tension}

\author{Bikash R. Dinda}
\email{bikashd18@gmail.com}
\affiliation{Department of Theoretical Physics, Tata Institute of Fundamental Research, Dr. Homi Bhabha Road, Navy Nagar, Colaba, Mumbai-400005, India.}

\begin{abstract}
We propose an analytical parametrization of the comoving distance and Hubble parameter to study the cosmic expansion history beyond the vanilla $\Lambda$CDM model. The parametrization is generalized enough to include the contribution of spatial curvature and to capture the higher redshift behaviors. With this parameterization, we study the late-time cosmic behavior and put constraints on the cosmological parameters like present values of Hubble parameter ($H_{0}$), matter-energy density parameter ($\Omega_{\text{m0}}$), spatial curvature energy density parameter ($\Omega_{\text{k0}}$) and baryonic matter-energy density parameter ($\Omega_{\text{b0}}$) using different combinations data like CMB (Cosmic microwave background), BAO (baryon acoustic oscillation), and SN (Pantheon sample for type Ia supernovae). We also rigorously study the Hubble tension in the framework of late time modification from the standard $\Lambda$CDM model. We find that the late time modification of the cosmic expansion can solve the Hubble tension between CMB $\&$ SHOES (local distance ladder observation for $H_{0}$), between CMB+BAO $\&$ SHOES and between CMB+SN $\&$ SHOES, but the late time modification can not solve the Hubble tension between CMB+BAO+SN and SHOES. That means CMB, BAO, and SN data combined put strong enough constraints on $H_{0}$ (even with varying $\Omega_{\text{k0}}$) and on other background cosmological parameters so that the addition of $H_{0}$ prior from SHOES (or from similar other local distance observations) can not significantly pull the $H_{0}$ value towards the corresponding SHOES value.
\end{abstract}


\maketitle
\date{\today}

\section{Introduction}

Since the discovery of the late time cosmic acceleration from supernovae type Ia observations in 1998 \citep{Riess:1998cb,Perlmutter:1998np,2011NatPh...7Q.833W}, this comic acceleration has been confirmed by many cosmological observations like Planck mission for cosmic microwave background (CMB) \citep{Ade:2015xua,Aghanim:2018eyx}, baryon acoustic oscillations (BAO) observations \citep{Delubac:2014aqe,Ata:2017dya}, cosmic chronometers measurement for Hubble parameter \citep{Pinho:2018unz} etc. After that, many theoretical models have been proposed to explain this acceleration. Two main broad classes of models are dark energy \citep{Copeland:2006wr,Tsujikawa:2013fta,Zlatev:1998tr,Steinhardt:1999nw,Caldwell:2005tm,Linder:2006sv,Tsujikawa:2010sc,Scherrer:2007pu,Dinda:2016ibo,Dinda:2017swh,Lonappan:2017lzt} and modified gravity models \citep{Clifton:2011jh,Hinterbichler:2011tt,deRham:2012az,deRham:2014zqa,DeFelice:2010aj,Nojiri:2010wj,Nojiri:2017ncd,Dinda:2017lpz,Dinda:2018eyt,Zhang:2020qkd}. In the dark energy models, the late time acceleration is caused by an exotic matter, called the dark energy, which has large negative pressure. This negative pressure introduces repulsive gravity which provides the cosmic acceleration within the framework of general relativity \citep{Copeland:2006wr,Tsujikawa:2013fta,Zlatev:1998tr,Steinhardt:1999nw}. In the second class of models i.e. in the modified gravity theories, the late time cosmic acceleration is caused by the modification to the general relativity without introducing any exotic matter \citep{Clifton:2011jh,Hinterbichler:2011tt,deRham:2012az}.

The simplest dark energy model is the $\Lambda$CDM model, where the late time cosmic acceleration is caused by a cosmological constant \citep{Ade:2015xua,Aghanim:2018eyx}. This model has been put to many observational tests \citep{Sahni:2014ooa,Lusso:2019akb,Demianski:2019vzl,Haslbauer:2020xaa,Riess:2016jrr,Bonvin:2016crt,Nesseris:2014mfa,DelPopolo:2016emo,Godlowski:2005tw,Vagnozzi:2018jhn,Rezaei:2020mrj} and in some cases, this model has some discrepencies with observational data to some extent \citep{Banerjee:2020bjq,Sahni:2014ooa,Lusso:2019akb,Demianski:2019vzl,Haslbauer:2020xaa}. Also, in the $\Lambda$CDM model framework, there is a discrepency in the $H_{0}$ value (present value of Hubble parameter) between the early Universe observations like CMB and the late time local distance ladder observations like SHOES \citep{Riess:2016jrr,Riess:2020fzl}. This is called the so called Hubble tension or $H_{0}$ tension \citep{DiValentino:2021izs,Krishnan:2021dyb,Vagnozzi:2019ezj,Visinelli:2019qqu,Dainotti:2021pqg}. Thus, these are the motivations to go beyond the $\Lambda$CDM model.

$\Lambda$CDM model along with other dark energy models like quintessence and k-essence have theoretical issues like cosmic coincidence and fine-tuning \citep{Zlatev:1998tr,Sahni:1999gb,Velten:2014nra,Malquarti:2003hn}. Despite these theoretical issues, there are some limitations to different dark energy models. For example, $\Lambda$CDM model possesses constant (fixed to $-1$) equation of state of the dark energy, quintessence models have evolving equation of state of the dark energy (eos, $w$) but restricted to the non-phantom ($w>-1$) regions \citep{Copeland:2006wr,Tsujikawa:2013fta,Zlatev:1998tr}. In the context of the Hubble tension, the quintessence models make it worse since this tension is worse for the non-phantom equation of state while the phantom equation of state improves this tension \citep{Banerjee:2020xcn}. To avoid these limitations, parametric models are useful. For example, wCDM \citep{2012ApJ...760...19P,Mortonson:2013zfa,Anselmi:2014nya} and Chevallier, Polarski and Linder (CPL) \citep{Chevallier:2000qy,Linder:2002et} parametrizations where equation of state of the dark energy, $w$ is constant and evolving respectively. Although wCDM model possess both phantom ($w<-1$) and non-phantom ($w>-1$) regions but it is restricted to a constant value \citep{2012ApJ...760...19P,Mortonson:2013zfa,Anselmi:2014nya}. However, CPL parameterization is widely used in cosmology because of its generalization that in this parameterization equation of state of dark energy is evolving with redshift and possesses both phantom ($w<-1$) and non-phantom ($w>-1$) regions \citep{Chevallier:2000qy,Linder:2002et}. There are other parametrizations of the equation of state of the evolving dark energy like Barboza and Alcaniz (BA) \citep{Barboza:2008rh} and generalized Chaplygin gas (GCG) \citep{Thakur:2012rp} parametrizations etc \citep{Dinda:2018ojk,Dinda:2018uwm}.

All the above-mentioned evolving dark energy parametrizations (CPL, BA, GCG, etc) are bi-dimensional (i.e. consists of two parameters or dark energy degrees of freedom $2$) and are (mostly) based on the Taylor series expansion \citep{Escamilla-Rivera:2016qwv}. Taylor series expansion has the limitation that it is accurate in the small argument limit, for example, in CPL parametrization, the equation of state of the dark energy is accurate when $(1-a) \ll 1$ and the errors in it increases when $(1-a) \sim 1$. So, it is still necessary to go beyond these parametrizations and we may need parametrization beyond dark energy degrees of freedom $2$ \citep{Colgain:2021pmf,Escamilla-Rivera:2016qwv,Zhai:2017vvt,Linden:2008mf,Liu:2008vy,Sendra_2012,Ferramacho_2010,Dinda:2019mev,Krishnan:2020obg,Kenworthy:2019qwq,Sapone:2014nna,dePutter:2007kf}.

With the above-mentioned motivations, we propose an analytic parametrization of the line of sight comoving distance and consequently for the Hubble parameter with some modification to the Taylor series expansion. This parametrization includes cosmic curvature and the radiation terms to see how these terms affect the late-time cosmic acceleration. Although the radiation term is not that effective to the late time cosmic acceleration, still inclusion of the non-zero radiation term is important since its effect increases at higher redshifts. However, the cosmic curvature term is important at late times. Because of its importance, in cosmology, many authors have used it in their analysis and put constraints on it from different cosmological observations \citep{Gao:2020irn,Liu:2020bzc,Yang:2020bpv,Liu:2020pfa,Chudaykin:2020ghx,Wang:2020dbt}. We also constrain the cosmic curvature from some important cosmological observations like CMB and BAO \citep{Vagnozzi:2020rcz,Vagnozzi:2020dfn}.

In literature, some authors (see \citep{DiValentino:2021izs} and references therein) try to solve the Hubble tension with the late time modification of the cosmic expansion by different dark energy models, modified gravity models, and some dark energy parametrizations. With this motivation, we also study this issue with our parametrization in detail. Recently, some authors like in \citep{Camarena:2021jlr,Efstathiou:2021ocp} have claimed that one should use $M$ (absolute magnitude of type Ia supernovae peak amplitude) prior instead of $H_{0}$ prior and try to solve the $M$ tension instead of the Hubble tension, while data like Pantheon sample for type Ia supernova peak magnitude is used in the data analysis. The reasons are mentioned in the main text and \citep{Camarena:2021jlr}. So, it is important to study the $M$ tension and we do so in Sec. \ref{sec-Mtension}.

This paper is organized as follows: in Sec. \ref{sec-loscmvdHubble}, we present analytical parametrizations to the line of sight comoving distance and the normalized Hubble parameter; in \ref{sec-dataanalysis}, we mention cosmological observational data we use in our analysis and we do data analysis using these data and put constraints on the parameters; in Sec. \ref{sec-H0tension}, we study the Hubble tension in details and check whether late time modification of the cosmic expansion can solve this issue; in Sec. \ref{sec-Mtension}, we check whether late time modification can also solve the $M$ tension; finally, in Sec. \ref{sec-conclusion}, we present our conclusion.

\section{Parametrization to the line of sight comoving distance and the Hubble parameter}
\label{sec-loscmvdHubble}

\subsection{Basics}

The line of sight comoving distance (denoted by $d_{c}$) is defined as $d_{c} = d_{H} D$ where $D$ is given by \citep{Hogg:1999ad}

\begin{equation}
D (z) = \int_{0}^{z} \frac{dz'}{E(z')},
\label{eq:defnD}
\end{equation}

\noindent
with $z$ (also $z'$) being the redshift. $E$ is the normalized Hubble parameter, defined as $E(z)=H(z)/H_{0}$, where $H$ is the Hubble parameter and $H_{0}$ is its present ($z=0$) value. $d_{H}$ is defined as $d_{H}=c/H_{0}$, where $c$ is the speed of light in vacuum. We call $D$ as the normalized line of sight comoving dustance.

The inverse normalized Hubble parameter ($E^{-1}$) can be computed from $D$ by taking derivative of Eq.~\eqref{eq:defnD} and hence $E(z)$ becomes

\begin{equation}
E(z) = \left[ \dfrac{dD(z)}{dz} \right]^{-1}.
\label{eq:EfromD}
\end{equation}

Form Eqs.~\eqref{eq:defnD} and~\eqref{eq:EfromD}, it is clear that if we make parametrization to $D$, we shall always get analytic form of both $D$ and $E$. On the other hand, if we make parametrization to $E$, we shall not always get an analytic expression for $D$. This is the reason that we shall first make parametrization to $D(z)$ and consequently, we shall get parametrization for $E(z)$.

\subsection{$E(z)$ and $D(z)$ in $\Lambda$CDM model}

We shall soon see that our parametrization is based on the correction to the flat $\Lambda$CDM model (we simplify call this as the $\Lambda$CDM model). So, we first mention the expression for the normalized Hubble parameter (denoted by $E_{\text{$\Lambda$CDM}}$) given by

\begin{equation}
E_{\text{$\Lambda$CDM}}(z) = \sqrt{\Omega _{\text{m0}}(1+z)^3+1-\Omega _{\text{m0}}},
\label{eq:LcdmE}
\end{equation}

\noindent
where $\Omega _{\text{m0}}$ is the present value of the matter-energy density parameter. Consequently, we get the normalized line of sight comoving distance (denoted by $D_{\text{$\Lambda $CDM}}$) for the $\Lambda$CDM model given by (obtained by putting Eq.~\eqref{eq:LcdmE} in Eq.~\eqref{eq:defnD})

\begin{eqnarray}
&& D_{\text{$\Lambda $CDM}} (z) = \frac{1}{\sqrt{1-\Omega _{\text{m0}}}} \Big{[} - \, _2F_1\left(\frac{1}{3},\frac{1}{2},\frac{4}{3},\frac{\Omega _{\text{m0}}}{\Omega _{\text{m0}}-1}\right) \nonumber\\
&& + (1+z) \, _2F_1\left(\frac{1}{3},\frac{1}{2},\frac{4}{3},\frac{(1+z)^3 \Omega _{\text{m0}}}{\Omega _{\text{m0}}-1}\right) \Big{]} ,
\label{eq:LcdmD}
\end{eqnarray}

\noindent
where $_2F_1$ is the hypergeometric function.

\subsection{The parametrization}

Eq.~\eqref{eq:EfromD} can be linearized in derivative by defining $E^{-1}=F$ given by

\begin{equation}
F = \dfrac{dD}{dz} = -a^2\dfrac{dD}{da},
\label{eq:FfromD}
\end{equation}

\noindent
where $a$ is the scale factor and it is related to the redshift given by $a=1/(1+z)$. Using the above equation, we can split $F$ into two terms corresponding to the splitting of $D$ in two terms respectively as follows:

\begin{eqnarray}
F &=& F_{\text{$\Lambda $CDM}}+F_{\text{extra}},
\label{eq:total_F} \\
D &=& D_{\text{$\Lambda $CDM}}+D_{\text{extra}},
\label{eq:total_D} \\
F_{\text{extra}} &=& \dfrac{dD_{\text{extra}}}{dz} = -a^2\dfrac{dD_{\text{extra}}}{da},
\label{eq:Split_F_extra}
\end{eqnarray}

\noindent
where $F_{\text{$\Lambda $CDM}}=E_{\text{$\Lambda $CDM}}^{-1}$. The subscript 'extra' is meant for the extra contribution to the $\Lambda $CDM one both for $F$ and $D$. Now our task is to parametrize the $D_{\text{extra}}$ term and consequently we will get analytical expression for $F_{\text{extra}}$ using Eq.~\eqref{eq:Split_F_extra}.

\subsubsection{Parametrization of $D_{\text{extra}}$}

We parametrize $D_{\text{extra}}$ with the Taylor series expansion (with respect to the scale factor, $a$) around the point $a=0$ with a modification given by $D_{\text{extra}} = R_0+a^m \sum_{i=0}^{n} P_{i} a^{i}$, where $R_0$ and $P_{i}$s are the parameters in this parametrization. Here, $n$ is an integer. At this stage, $m$ can have any positive value i.e. $m \geq 0$. The reason to choose this kind of parametrization is as follows: first of all, Eq.~\eqref{eq:defnD} suggests that the value of $D$ at present is zero i.e. $D(z=0)=0$ and its value continuously increasing with increasing redshift provided $E(z)>0$ (which is the case for the standard cosmological scenario). Plus, we know that after a certain high redshift, $D(z)$ approximately approaches a constant value provided $E(z)$ is also a continuously increasing function with increasing $z$ (which is also the case for the standard cosmological scenario). We impose this fact through this parametrization, by the consideration that $D_{\text{extra}}$ approaches a constant value, $R_{0}$ for $a \ll 1$. And this is possible for $m \geq 0$. So, $D$ has the limit given by

\begin{eqnarray}
D & \sim & R_0 - \frac{\, _2F_1\left(\frac{1}{3},\frac{1}{2};\frac{4}{3};\frac{\Omega _{\text{m0}}}{\Omega _{\text{m0}}-1}\right)}{\sqrt{1-\Omega _{\text{m0}}}} \nonumber\\
&& + \frac{\Gamma \left(\frac{1}{6}\right) \Gamma \left(\frac{4}{3}\right)}{\sqrt{\pi } \sqrt[6]{1-\Omega _{\text{m0}}} \sqrt[3]{\Omega _{\text{m0}}}} + \mathcal{O} \left( a^{ \epsilon } \right) ,
\label{eq:appxDloweras}
\end{eqnarray}

\noindent
with $\epsilon>0$. The second and the third terms (i.e. the terms without $R_0$) in the right hand side of the above equation is $a \ll 1$ limit to $D_{\text{$\Lambda $CDM}}$ in Eq.~\eqref{eq:LcdmD}. For a special case, $\Omega _{\text{m0}}=0.3$, the above equation becomes $D (\Omega _{\text{m0}}=0.3) \sim 3.30508+R_0 + \mathcal{O} \left( a^{ \epsilon } \right) $.

Now, we rewrite the series expansion by defining $n=d+1$ and $P_{i+1}=Q_{i}$ $\forall \hspace{0.1 cm} i \in (1,d)$ given by

\begin{equation}
D_{\text{extra}} = R_0+a^m \left( P_0+P_1 a + \sum_{i=1}^{d} Q_{i} a^{1+i} \right),
\label{eq:prmtrznDextra}
\end{equation}

At present ($a=1$), $D$ should be zero. This immediately gives

\begin{equation}
R_0 = -P_0-P_1-\sum _{i=1}^d Q_i.
\label{eq:R0}
\end{equation}

At this stage, no further restrictions or constraints are required, if we are considering the parametrization for $D(z)$ only. But this is not complete yet. We shall have other constraints, while we derive the parametrization for $E(z)$ from $D(z)$. We shall see this next.

\subsubsection{Derived parametrization of $F_{\text{extra}}$}

Using Eq.~\eqref{eq:Split_F_extra}, we get derived parametrization for $F_{\text{extra}}$ from Eq.~\eqref{eq:prmtrznDextra} given by

\begin{eqnarray}
F_{\text{extra}} = & - a^{-(m+1)} & \Big{[} mP_0+(m+1)P_1a \nonumber\\
&& + \sum_{i=1}^{d} (m+1+i) Q_{i} a^{1+i} \Big{]}.
\label{eq:Fextra}
\end{eqnarray}

From the above expression, we can see that the term corresponding to the lowest power in $a$ is given by

\begin{equation}
\text{Term} \hspace{0.1 cm} (\text{lowest power in a}) = - m P_0 a^{m+1} .
\label{eq:FextraLowesta}
\end{equation}

At sufficiently higher redshifts (after the radiation-dominated era), the Universe is dominated by matter. We want to impose this condition in $E^{2}$ in Eq.~\eqref{eq:Fextra}. To do so, for the time being, we validate our parametrization from the present time to the matter-dominated era. So, at this moment, we are not considering radiation or any early Universe contributions (later we will show how to modify the parametrization for the early Universe contribution). That means we want to make the expression of $F$ in Eq.~\eqref{eq:total_F} such that at $z \gg 1$ (i.e. $a \ll 1$), $F$ behaves as in the matter-dominated era. To do this, we first check how $F$ behaves at the small $a$ limit for some special cases given below

\begin{eqnarray}
F_{\text{Matter+Curvature}} & \sim & \frac{a^{ \frac{3}{2} }}{\sqrt{\Omega _{\text{m0}}}}-\frac{a^{ \frac{5}{2} } \Omega _{\text{k0}}}{2 \Omega _{\text{m0}}^{ \frac{3}{2} }}+ \mathcal{O} \left(a^{ \frac{7}{2} }\right),
\label{eq:FmatterCurvLowera} \\
F_{\text{$\Lambda $CDM+Curvature}} & \sim & \frac{a^{ \frac{3}{2} }}{\sqrt{\Omega _{\text{m0}}}}-\frac{a^{ \frac{5}{2} } \Omega _{\text{k0}}}{2 \Omega _{\text{m0}}^{ \frac{3}{2} }}+ \mathcal{O} \left(a^{ \frac{7}{2} }\right),
\label{eq:FoLCDMlowera} \\
F_{\text{$\Lambda $CDM}} & \sim & \frac{a^{ \frac{3}{2} }}{\sqrt{\Omega _{\text{m0}}}}+ \mathcal{O} \left(a^{ \frac{7}{2} }\right),
\label{eq:LcdmFLowera}
\end{eqnarray}

\noindent
where, in Eq.~\eqref{eq:FmatterCurvLowera}, we considered matter and curvature contributions to total $E^2$. In Eq.~\eqref{eq:FoLCDMlowera}, we considered matter, cosmological constant and curvature contributions to total $E^2$. This is the non-flat $\Lambda$CDM model and we call this as o$\Lambda$CDM model. In Eq.~\eqref{eq:LcdmFLowera}, we considered matter and cosmological constant only to the total $E^2$. This is the flat $\Lambda$CDM model. $\Omega_{k0}$ is the present value of the curvature energy density parameter. It is defined as $\Omega_{k0}=-Kc^{2}/H_{0}^{2}$ with $K$ being the curvature of the spacetime. $K<0$, $K=0$, and $K>0$ correspond to the open, flat, and closed Universe respectively.

So, from the above small $a$ limit in Eq.~\eqref{eq:LcdmFLowera}, it is clear that the $\Lambda$CDM model (which is a subset in our parametrization) already gives required behavior at matter-dominated era. Another way to see this is that, the extra term, $F_{\text{extra}}$ should be such that, it becomes dominated after the early matter-dominated era. To get this behaviour, the required conditions on $m$ becomes (by comparing Eqs.~\eqref{eq:FextraLowesta} and~\eqref{eq:LcdmFLowera}) $m+1>3/2$ i.e. $m>1/2$. Note that, this condition does not violate the previous condition $m>0$.

The condition, $m>1/2$ is not complete yet, because we want to include the curvature term. To have this in our parametrization, we impose: $\text{Term} \hspace{0.1 cm} (\text{lowest power in a}) =$ Curvature term i.e. the term with $a^{5/2}$. This is just making the extra correction only in the late time era with the inclusion of curvature terms. So, $- m P_0 a^{m+1} = -\frac{a^{5/2} \Omega _{\text{k0}}}{2 \Omega _{\text{m0}}^{3/2}}$. So, we get

\begin{eqnarray}
m &=& \frac{3}{2},
\label{eq:mwrtn} \\
P_{0} &=& \frac{\Omega _{\text{k0}}}{3 \Omega _{\text{m0}}^{3/2}} .
\label{eq:Pnplus1}
\end{eqnarray}

\noindent
Note that, now we have got a fixed value of $m$ which does not violate the previous condition $m>1/2$. Also, note that the above condition is equivalent to

\begin{equation}
E^2 \sim \Omega_{\text{m0}} a^{-3} + \Omega_{\text{k0}} a^{-2} + \mathcal{O} \left(a^{-1}\right) ,
\label{eq:EsqrLowera}
\end{equation}

\noindent
where $E$ is (computed from Eq.~\eqref{eq:total_F}) given by

\begin{equation}
E = \frac{1}{F_{\text{$\Lambda $CDM}}+F_{\text{extra}}} .
\label{eq:E}
\end{equation}

We are left with one important constraint that at present $E$ should be unity i.e. $E(a=1)=1$. This can be rewritten as $F_{\text{extra}}(a=1)=0$, because $F_{\text{$\Lambda $CDM}}(a=1)=1=E_{\text{$\Lambda $CDM}}(a=1)$. Putting this constraint in equation Eq.~\eqref{eq:Fextra}, we get

\begin{equation}
P_{1} = - \frac{1}{5} \left[ \frac{\Omega _{\text{k0}}}{\Omega _{\text{m0}}^{3/2}} + \sum _{i=1}^d (2 i+5) Q_i \right].
\label{eq:P1noearly}
\end{equation}

So, in our parametrization, the free parameters are $Q_{1}$, $Q_{2}$, ..., $Q_{d}$. These can be considered as the dark energy parameters with the dark energy degrees of freedom $d$. For example, $d=0$ case is the minimal case, where there are no free parameters and the dark energy degree of freedom is 0. Note that, in our parametrization, the $\Lambda$CDM model is the special case when $d=0$ and $\Omega_{\text{k0}}=0$. For the case of $d=1$, there is only one free parameter $Q_{1}$ and the dark energy degrees of freedom is 1. For the case of $d=2$, there are two free parameters $Q_{1}$ and $Q_{2}$ and the dark energy degrees of freedom is 2 and so on.

Another point to notice that, at first order in $z$, we get $D \simeq z$ i.e. $d_{c} \simeq d_{H}z$. This further implies $v \simeq H_{0}d_{c}$, where $v=cz$ is the recessional velocity. This is the Hubble's law. So, our parametrization is consistent with the Hubble's law.

\subsection{Modification to include the early Universe contribution}

So far, we have considered that $E$ behaves like matter-dominated at higher redshifts instead of radiation-dominated or others i.e. the parametrization of $E$ in Eq.~\eqref{eq:E} is valid for the late time and matter-dominated eras only. Now, we want to discuss how to include any early universe correction, for example, the inclusion of the radiation-dominated era. In our parametrization, we can simply do that by adding the early Universe correction to $E^{2}$ without correcting the parametrization to $D$. This is because, in the standard cosmology, $E^{-1}$ is a monotonically decreasing function with increasing $z$. This further ensures that (from Eq.~\eqref{eq:defnD}), after certain redshift, $D$ approaches nearly a constant value (the zeroth-order term in $a$ in the right hand side of Eq.~\eqref{eq:appxDloweras}). This is a good assumption that this saturation already happens in the matter-dominated era and any early Universe correction (before the matter-dominated era) like the radiation term correction does not change the values of $D$ significantly. So, we can express total $E^{2}$ (denoted by $E^{2}_{T}$) as

\begin{equation}
E^{2}_{T}(z) = E^{2}(z)+E^{2}_{\text{early}}(z),
\label{eq:TotalEsqr}
\end{equation}

\noindent
where $E^{2}_{\text{early}}$ is the early Universe correction.

At present ($z=0$), $E_{T}$ should be unity. This immediately fixes $P_{1}$ (from Eqs.~\eqref{eq:E} and~\eqref{eq:TotalEsqr}) given by

\begin{equation}
P_{1} = - \frac{1}{5} \left[ \left( \frac{2}{\sqrt{1-A_0}} -2 \right) + \frac{\Omega _{\text{k0}}}{\Omega _{\text{m0}}^{3/2}} + \sum _{i=1}^d (2 i+5) Q_i \right].
\label{eq:earlyP1}
\end{equation}

\noindent
where $A_{0}=E^{2}_{\text{early}}(z=0)$. Now, we can see that we get back Eq.~\eqref{eq:P1noearly}, if we put $A_{0}=0$ in Eq.~\eqref{eq:earlyP1}.

As mentioned before, the total $D$ (denoted by $D_{T}$) is assumed to be the same as in Eq.~\eqref{eq:total_D} i.e.

\begin{equation}
D_{T} \simeq D.
\label{eq:TotalD}
\end{equation}

The above assumption is valid for $A_{0} \ll 1$ and this is the case since this is the early Universe correction.

\subsubsection{Validity of the assumption in Eq.~\eqref{eq:TotalD}}

\begin{figure}
\includegraphics[width=0.45\textwidth]{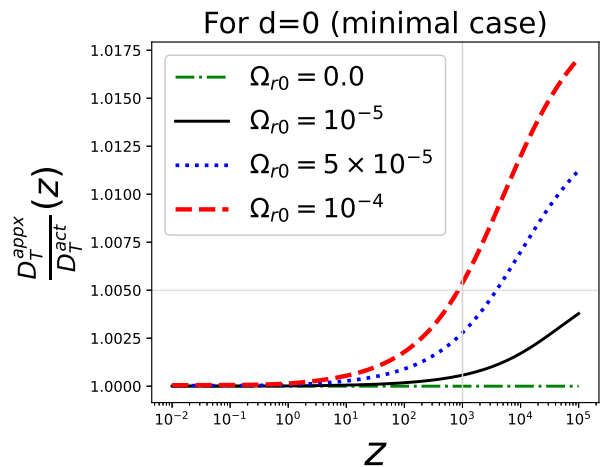}
\caption{Comparison of the approximated $D_{T}$ (computed from Eq.~\eqref{eq:TotalD} and denoted by $D_{T}^{appx}$) with the total accurate one (computed by putting Eq.~\eqref{eq:TotalEsqr} in Eq.~\eqref{eq:defnD} and denoted by $D_{T}^{act}$). Here, we have considered $d=0$ which is the minimal case, where there is no dark energy degrees of freedom. Here, $E^{2}_{\text{early}}(z)=\Omega_{\text{r0}}(1+z)^{4}$. The parameter values, we fix, are given by $\Omega_{\text{m0}}=0.3$ and $\Omega_{\text{k0}}=0$. We consider four choices of $\Omega_{\text{r0}}$ values and these are $\Omega_{\text{r0}}=0.0$ (dashed-dotted green line), $\Omega_{\text{r0}}=10^{-5}$ (solid black line), $\Omega_{\text{r0}}=5 \times 10^{-5}$ (dotted blue line) and $\Omega_{\text{r0}}=10^{-4}$ (dashed red line). Note that, $D_{T}^{appx}$ can be seen as $D_{T}^{act}$ with $\Omega_{\text{r0}}=0$.}
\label{fig:Dappx}
\end{figure}

Let us shortly discuss the validity of the assumption in Eq.~\eqref{eq:TotalD} by considering early Universe correction is the radiation correction only i.e. $E^{2}_{\text{early}}(z)=\Omega_{\text{r0}}(1+z)^{4}$, where $\Omega_{\text{r0}}$ is the present value of the radiation energy density parameter. In Fig.~\ref{fig:Dappx}, we compare the approximated total $D$ (denoted by $D_{T}^{appx}$; using Eq.~\eqref{eq:TotalD}) with the actual total $D$ (denoted by $D_{T}^{act}$; putting Eq.~\eqref{eq:TotalEsqr} in Eq.~\eqref{eq:defnD}) for four choices of $\Omega_{\text{r0}}$ values mentioned in the figure. In this figure, we have considered $d=0$ (minimal case). We see that the accuracy is at percentage level (up to $0.5\%$ and $1.75\%$ at $z=10^{3}$ and $z=10^{5}$ respectively). Note that we have checked this fact for $d=0$ case only, but one can easily check that, for other values of $d$, this fact is true.

In the next sections, we will not consider any early Universe corrections, since our next discussions do not include higher redshifts like $z>1100$. So, in the next sections, we fix $E^{2}_{\text{early}}(z)=0$ (or $A_{0}=0$).

\subsection{Some other derived quantities}

From the line of sight comoving distance, $d_c(z)$, we can compute the transverse comoving distance, $d_M(z)$ given by \citep{Hogg:1999ad}

\begin{equation}
    d_M = \begin{cases}
    \frac{d_{H}}{\sqrt{\Omega_{\text{k0}}}} \sinh \left( \sqrt{\Omega_{\text{k0}}} D \right), & \mbox{if } \Omega_{\text{k0}}>0, \\
    d_H D, & \mbox{if } \Omega_{\text{k0}} = 0, \\
    \frac{d_{H}}{\sqrt{|\Omega_{\text{k0}}|}} \sin \left( \sqrt{|\Omega_{\text{k0}}|} D \right), & \mbox{if } \Omega_{\text{k0}}<0 . \\
    \end{cases}
    \label{eq:trnscov}
\end{equation}

The luminosity distance and the angular diameter distance are given by

\begin{eqnarray}
d_L &=& \left( 1+z \right) d_M,
\label{eq:Luminosity} \\
d_A &=& \frac{d_M}{1+z},
\label{eq:angular}
\end{eqnarray}

\noindent
respectively. The deceleration parameter can be written as

\begin{equation}
q = \left( 1+z \right) \frac{E'}{E} -1,
\label{eq:q}
\end{equation}

\noindent
where $'$ represents the derivative with respect to the redshift. The equation of the state of the dark energy can be written as

\begin{equation}
w = - \left( \frac{1}{3} \right) \frac{ \Omega _{\text{k0}} (1+z)^2 + 2(1+z)E E' - 3 E^2 }{ \Omega _{\text{m0}} (1+z)^3 + \Omega _{\text{k0}} (1+z)^2 -E^2 }.
\label{eq:w}
\end{equation}

\subsection{Summary}

We now present the main equations of the parametrization as a summary given by

\begin{eqnarray}
&& D = D_{\text{$\Lambda $CDM}} + \frac{\left[(5-3 a) a^{3/2}-2\right] \Omega _{\text{k0}}}{15 \Omega _{\text{m0}}^{3/2}} \nonumber\\
&& + \sum _{i=1}^d \frac{1}{5} \left[-\left(a^{5/2} (2 i+5)\right)+5 a^{i+\frac{5}{2}}+2 i\right] Q_i
\end{eqnarray}

\begin{eqnarray}
F &=& D_{\text{$\Lambda $CDM}} + \frac{a^{5/2}}{2} \Big{[} \frac{(a-1) \Omega _{\text{k0}}}{\Omega _{\text{m0}}^{3/2}} \nonumber\\
&& + \sum _{i=1}^d a (2 i+5) \left(1-a^i\right) Q_i \Big{]}.
\end{eqnarray}

\noindent
We have mainly summarised the final expressions for $D$ and $F$ respectively. The other expressions are straightforward.

\begin{figure*}
\includegraphics[width=0.9\textwidth]{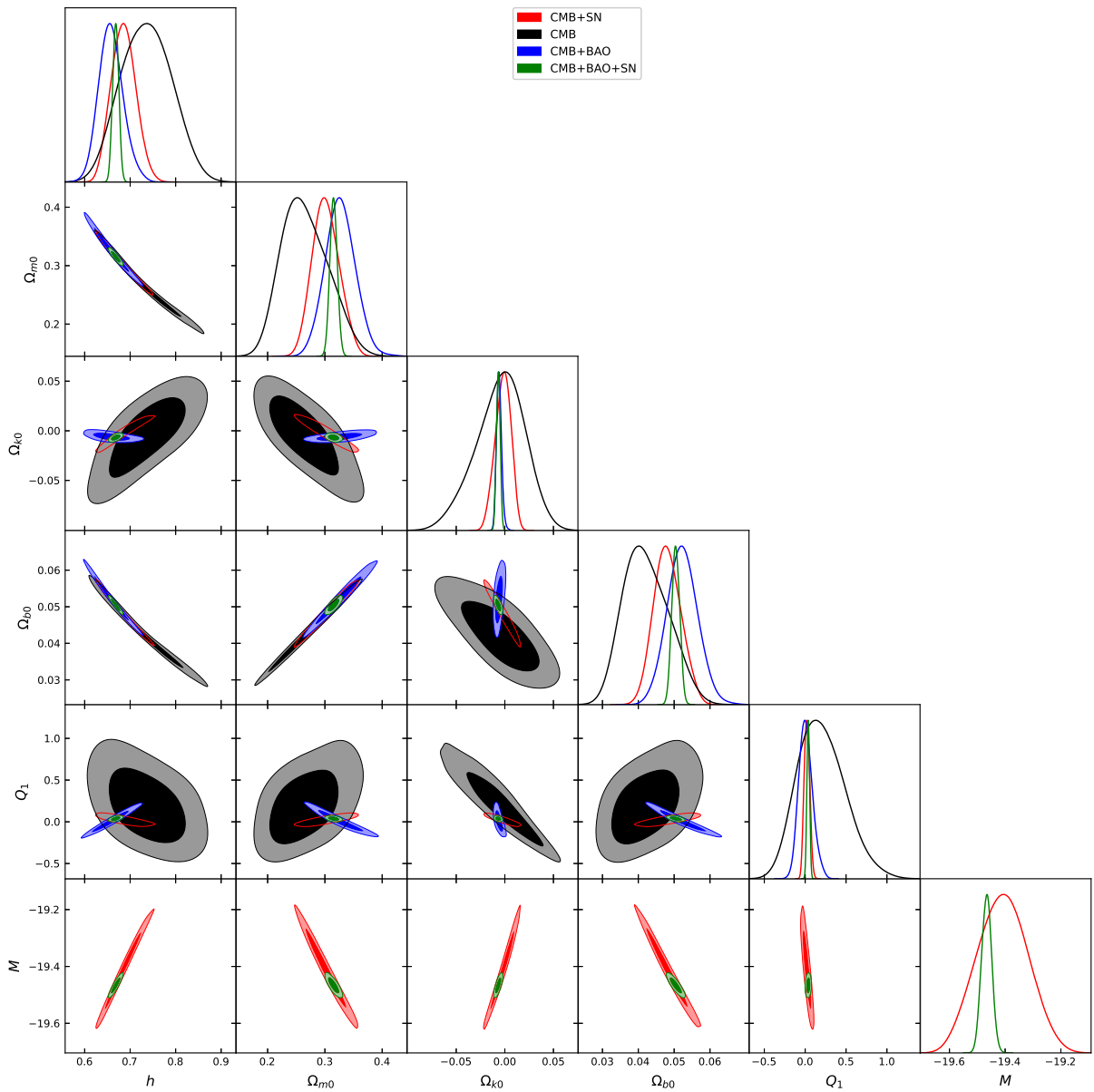}
\caption{Triangle plot to show constraints on model parameters for $d=1$ case for four combinations of data sets CMB (black), CMB+BAO (blue), CMB+SN (red) and CMB+BAO+SN (green). In this case, there is only one dark energy parameter, $Q_{1}$ i.e. dark energy degrees of freedom is 1.}
\label{fig:trngld1}
\end{figure*}

\begin{figure*}
\includegraphics[width=0.95\textwidth]{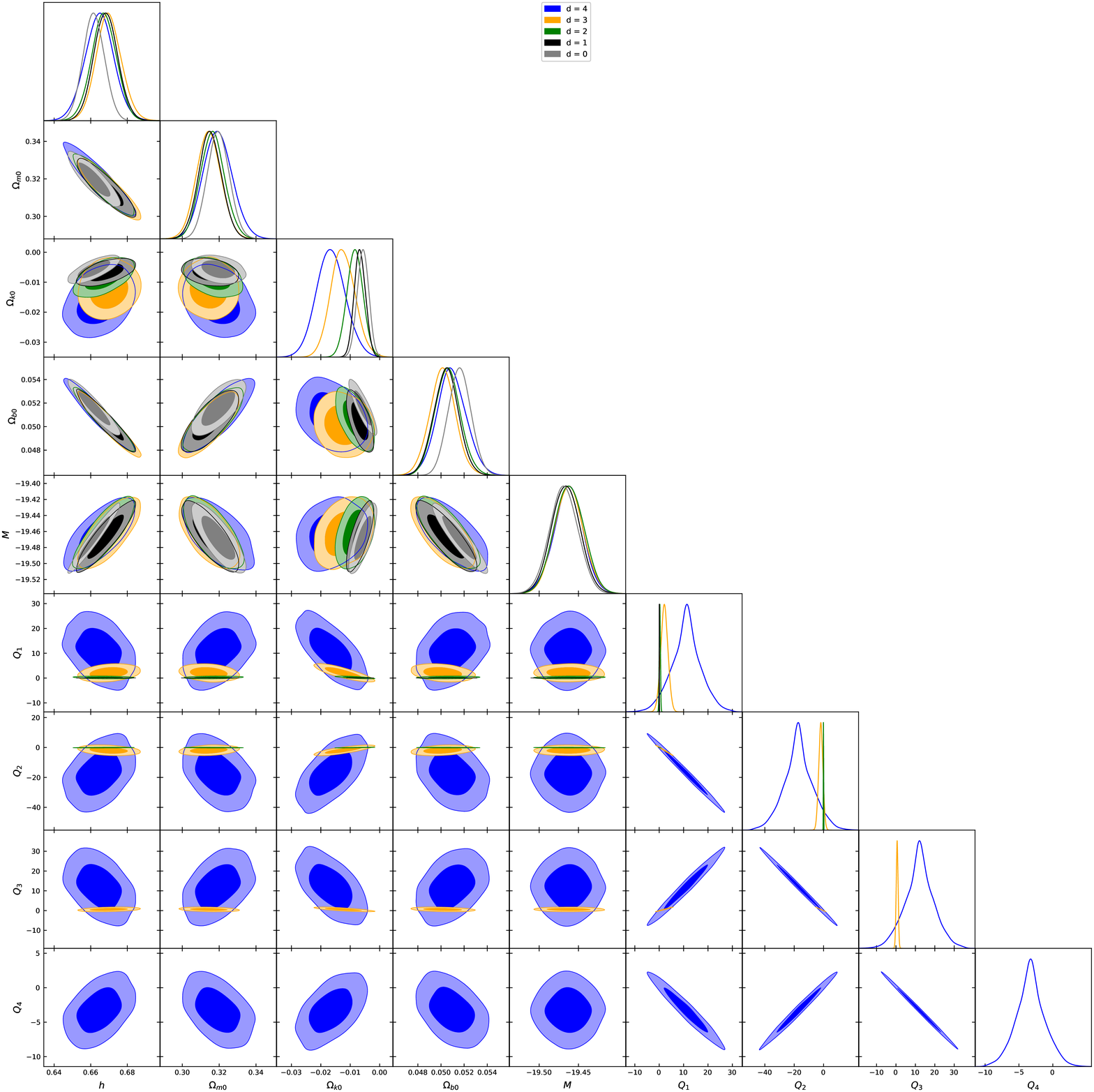}
\caption{Triangle plot for different parametrizations corresponding to different $d$ to show how contour areas (covariances) among cosmological parameters change with changing $d$. Here, we consider CMB+BAO+SN combination of data sets.}
\label{fig:trnglndffrntn}
\end{figure*}

\section{Data analysis and model comparison}
\label{sec-dataanalysis}

We mainly consider three kinds of cosmological data which are listed below:

\begin{itemize}
\item
(1) We use the Planck 2018 results for TT,TE,EE+lowl+lowE+lensing from the cosmic microwave background (CMB) observation for the base $\Lambda$CDM model (with and without cosmic curvature) as prior \citep{Aghanim:2018eyx,Zhai:2018vmm,Chen:2018dbv}. We denote this as 'CMB'.
\item
(2) We consider the BAO measurements from different surveys. For this, we closely follow \citep{Alam:2020sor} (see references therein). We exclude the measurement of eBOSS (the extended baryon oscillation spectroscopic survey ) emission-line galaxies (ELGs) at $z=0.85$ from \citep{Alam:2020sor} because of asymmetric standard deviation. We denote this as 'BAO'.
\item
(3) We also consider the Pantheon data for supernovae type Ia observation \citep{Scolnic:2017caz}. We denote this as 'SN'.
\end{itemize}

In this section, we constrain our model parameters along with the cosmological parameters like $h$, $\Omega_{\text{b0}}$ and $M$ (including the cosmological parameters, $\Omega_{\text{m0}}$ and $\Omega_{\text{k0}}$) with the cosmological data, mentioned above. Here, $h$ is defined as $H_{0}=100 \hspace{0.1 cm} h$ km s$^{-1}$ Mpc$^{-1}$. $\Omega_{\text{b0}}$ is the present value of the baryonic matter energy density parameter. It arises both in the CMB distance prior data and BAO data. Here, $M$ is the peak absolute magnitude of type Ia supernovae. $M$ arises only when SN data is used \citep{Scolnic:2017caz}.

To do parameter estimation, we consider flat priors on the parameters given by

\begin{eqnarray}
0.4 \leq & h & \leq 1, \nonumber\\
0 \leq & \Omega_{\text{m0}} & \leq 1, \nonumber\\
-0.5 \leq & \Omega_{\text{k0}} & \leq 0.5, \nonumber\\
0 \leq & \Omega_{\text{b0}} & \leq 0.2, \nonumber\\
-21 \leq & M & \leq -18, \nonumber \\
-100 \leq & Q_{i} & \leq 100. \nonumber
\end{eqnarray}

\begin{table*}
\begin{center}
\begin{tabular}{ |c|c|c|c|c|  }
\hline
 & CMB  & CMB+BAO  & CMB+SN  & CMB+BAO+SN \\
\hline
$h$ & $0.736\pm0.054$  & $0.658^{+0.024}_{-0.028}$ & $0.685\pm0.027$ & $0.6683\pm0.0067$ \\
\hline
$\Omega_{\text{m0}}$ & $0.265^{+0.035}_{-0.046}$ & $0.327\pm0.026$ & $0.302^{+0.022}_{-0.025}$ & $0.3152\pm0.0064$ \\
\hline
$\Omega_{\text{k0}}$ & $-0.005^{+0.028}_{-0.022}$  & $-0.0059^{+0.0023}_{-0.0028}$  & $-0.0019^{+0.0089}_{-0.0074}$  & $-0.0067\pm0.0019$ \\
\hline
$\Omega_{\text{b0}}$ & $0.0422^{+0.0056}_{-0.0072}$  & $0.0523\pm0.0043$  & $0.0481^{+0.0036}_{-0.0041}$  & $0.0504\pm0.0011$ \\
\hline
$Q_1$ & $0.20^{+0.26}_{-0.32}$  & $0.007^{+0.075}_{-0.092}$  & $0.022\pm0.033$  & $0.038\pm0.016$ \\
\hline
$M$ & -  & -  & $-19.409\pm0.090$  & $-19.466\pm0.018$ \\
\hline
\end{tabular}
\end{center}
\caption{ $1\sigma$ ranges of the parameters for the case of $d=1$ for CMB, CMB+BAO, CMB+SN, and CMB+BAO+SN combinations of data sets.}
\label{table:d1case}
\end{table*}

In Fig.~\ref{fig:trngld1}, we constrain model parameters for $d=1$ case for four combinations of data sets given by CMB (black), CMB+BAO (blue), CMB+SN (red), and CMB+BAO+SN (green). This is the case where the dark energy degree of freedom is $1$ and there is only one dark energy parameter which is $Q_1$. We list the $1\sigma$ ranges of the parameters in Table~\ref{table:d1case}. We see that 1$\sigma$ confidence regions are relatively larger for CMB-only data. When combined with one of the BAO and SN data, the confidence regions become tighter. When CMB, BAO, and SN data are combined altogether, the confidence regions become significantly tighter. This is true both for cosmological parameters ($h$, $\Omega_{m0}$, $\Omega_{k0}$, $\Omega_{b0}$, and $M$) and the dark energy degrees of freedom related parameter, $Q_{1}$. Note that, we have not shown the contours for other cases for different combinations of data sets (except for CMB+BAO+SN). But one can check that a similar conclusion can be drawn for the other values of $d$. We only show the results for CMB+BAO+SN data for different values of $d$ in the next figure (to avoid showing many plots and tables).

\begin{table*}
\begin{center}
\begin{tabular}{ |c|c|c|c|c|c|  }
\hline
 & $d=0$  & $d=1$  & $d=2$  & $d=3$ & $d=4$ \\
\hline
$h$ & $0.6616\pm0.0058$  & $0.6683\pm0.0066$ & $0.6672\pm0.0068$ & $0.6698\pm0.0069$ & $0.6645\pm0.0083$ \\
\hline
$\Omega_{\text{m0}}$ & $0.3197\pm0.0061$ & $0.3151\pm0.0063$ & $0.3165\pm0.0066$ & $0.3144\pm0.0067$ & $0.3193\pm0.0080$ \\
\hline
$\Omega_{\text{k0}}$ & $-0.0057\pm0.0019$  & $-0.0067\pm0.0019$  & $-0.0084\pm0.0027$  & $-0.0125\pm0.0041$ & $-0.0165\pm0.0049$ \\
\hline
$\Omega_{\text{b0}}$ & $0.05163\pm0.00096$  & $0.0504\pm0.0011$  & $0.0506\pm0.0011$  & $0.0501\pm0.0011$ & $0.0510\pm0.0013$ \\
\hline
$M$ & $-19.469\pm0.018$  & $-19.466\pm0.018$  & $-19.462\pm0.019$  & $-19.462\pm0.018$ & $-19.463\pm0.019$ \\
\hline
$Q_1$ & -  & $0.037\pm0.016$  & $0.25\pm0.23$  & $2.1\pm1.5$ & $11.0\pm5.8$ \\
\hline
$Q_2$ & -  & -  & $-0.097\pm0.11$  & $-1.9\pm1.4$ & $-17.0^{+8.3}_{-10}$ \\
\hline
$Q_3$ & -  & -  & -  & $0.60\pm0.49$ & $12.1\pm7.2$ \\
\hline
$Q_4$ & -  & -  & -  & - & $-3.3\pm2.0$ \\
\hline
\end{tabular}
\end{center}
\caption{ $1\sigma$ ranges of the parameters for the CMB+BAO+SN combination of data sets for $d=0,1,2,3$ and $4$. }
\label{table:parametervalues}
\end{table*}

In Fig.~\ref{fig:trnglndffrntn}, we consider all three data together i.e. the combination CMB+BAO+SN combination of data sets and vary $d$ (number of dark energy degrees of freedom) to show how contour-areas increase with increasing degrees of freedom. Gray, black, green, orange, and blue contour regions are for $d=0,1,2,3$ and $4$ respectively. We list $1\sigma$ ranges of the parameters in Table~\ref{table:parametervalues} for CMB+BAO+SN data for different values of $d$. We can see that the variances of each parameter and the contour areas of each pair of the parameters increase with increasing $d$. However, an important point to notice here is that this increment is significant for the dark energy degrees of freedom related parameters i.e. for $Q_i$s, whereas this increment is not very significant for the cosmological parameters except the $\Omega_{\text{k0}}$ parameter. The interesting point to notice that the standard deviation in the $\Omega_{\text{k0}}$ parameter increases significantly with increasing $d$ for lower values of $d$, but after a certain values of $d$ (around $d=3$ or $d=4$), this increment gradually becomes insignificant. This means the $\Omega_{\text{k0}}$ parameter is strongly correlated to the dark energy parameters when the dark energy degrees of freedom is smaller, but for larger dark energy degrees of freedom, this correlation becomes weaker. It is, thus, important to include spatial curvature terms in the cosmological data analysis. Other cosmological parameters are weakly correlated to the dark energy parameters. We shall later see that this fact will be reflected in the next sections when we talk about the Hubble tension or M tension.

\begin{table}
\begin{center}
\begin{tabular}{ |c|c|c|c|c|  }
\hline
model & $\ln \mathcal{Z}$  & $\Delta _1$  & AIC  & $\Delta _2$ \\
\hline
o$\Lambda$CDM & $552.1$  & $0$  & $1085.86$  & $0$ \\
\hline
owCDM & $551.9$  & $-0.2$  & $1081.3$  & $-4.56$ \\
\hline
oCPL & $548.5$  & $-3.6$  & $1083.51$  & $-2.35$ \\
\hline
$d=0$ & $552.4$  & $0.3$  & $1085.32$  & $-0.54$ \\
\hline
$d=1$ & $549.7$  & $-2.4$  & $1082.12$  & $-3.74$ \\
\hline
$d=2$ & $546.6$  & $-5.5$  & $1082.98$  & $-2.88$ \\
\hline
$d=3$ & $544.9$  & $-7.2$  & $1135.93$  & $50.07$ \\
\hline
$d=4$ & $539.1$  & $-13.0$  & $1595.82$  & $509.96$ \\
\hline
\end{tabular}
\end{center}
\caption{ Model comparison by $\ln \mathcal{Z}$ and AIC. Here we consider CMB+BAO+SN combinations of dataset. $\Delta _1=\ln \mathcal{Z}-\ln \mathcal{Z}(o\Lambda\text{CDM})$ and similarly $\Delta _2=\text{AIC}-\text{AIC}(o\Lambda\text{CDM})$.}
\label{table:modelcomparison}
\end{table}

Now, we briefly discuss the comparison of our parametrization to some standard cosmological models. For the standard cosmological models, we consider three types of models given by o$\Lambda$CDM, o$w$CDM, and oCPL. The o$\Lambda$CDM, o$w$CDM, and oCPL model correspond to the dark energy equation state $-1$, constant, and evolving respectively (for the details of these parameters, see \citep{Chevallier:2000qy,Linder:2002et}). The prescript 'o' represents the presence of the spatial curvature terms in these models. With these models, we compare our parametrizations (for the cases from $d=0$ to $d=4$) in two ways. One is by comparing the $\ln \mathcal{Z}$ values, where $\mathcal{Z}$ is the Bayesian posterior probability distribution. Another way is by comparing the Akaike Information Criterion (AIC). AIC is defined as $\text{AIC}=\chi^2_{\text{min}}+2\kappa$, where $\chi^2_{\text{min}}$ is the chi-square value corresponding to the best fit values of the parameters obtained from the data analysis. $\kappa$ is the total number of parameters for a particular parametrization or model. We list the $\ln \mathcal{Z}$ and AIC values in Table~\ref{table:modelcomparison}. We also compare the values of $\ln \mathcal{Z}$ and AIC with the o$\Lambda$CDM model for a particular parametrization by defining $\Delta_1$ and $\Delta_2$ such that $\Delta _1=\ln \mathcal{Z}-\ln \mathcal{Z}(o\Lambda\text{CDM})$ and $\Delta _2=\text{AIC}-\text{AIC}(o\Lambda\text{CDM})$ respectively. If a model has a positive $\Delta _1$ value it is better compared to o$\Lambda$CDM and it is significantly better if $\Delta _1>5$. On the other hand, if a model has a negative $\Delta _2$ value it is better compared to o$\Lambda$CDM. Comparing both the methods, we can see that our parametrizations with $0 \leq d \leq 2$ are equally competitive models compared to the o$\Lambda$CDM, o$w$CDM, and oCPL models.

\begin{figure*}[]
\begin{center}
\resizebox{250pt}{140pt}{\includegraphics{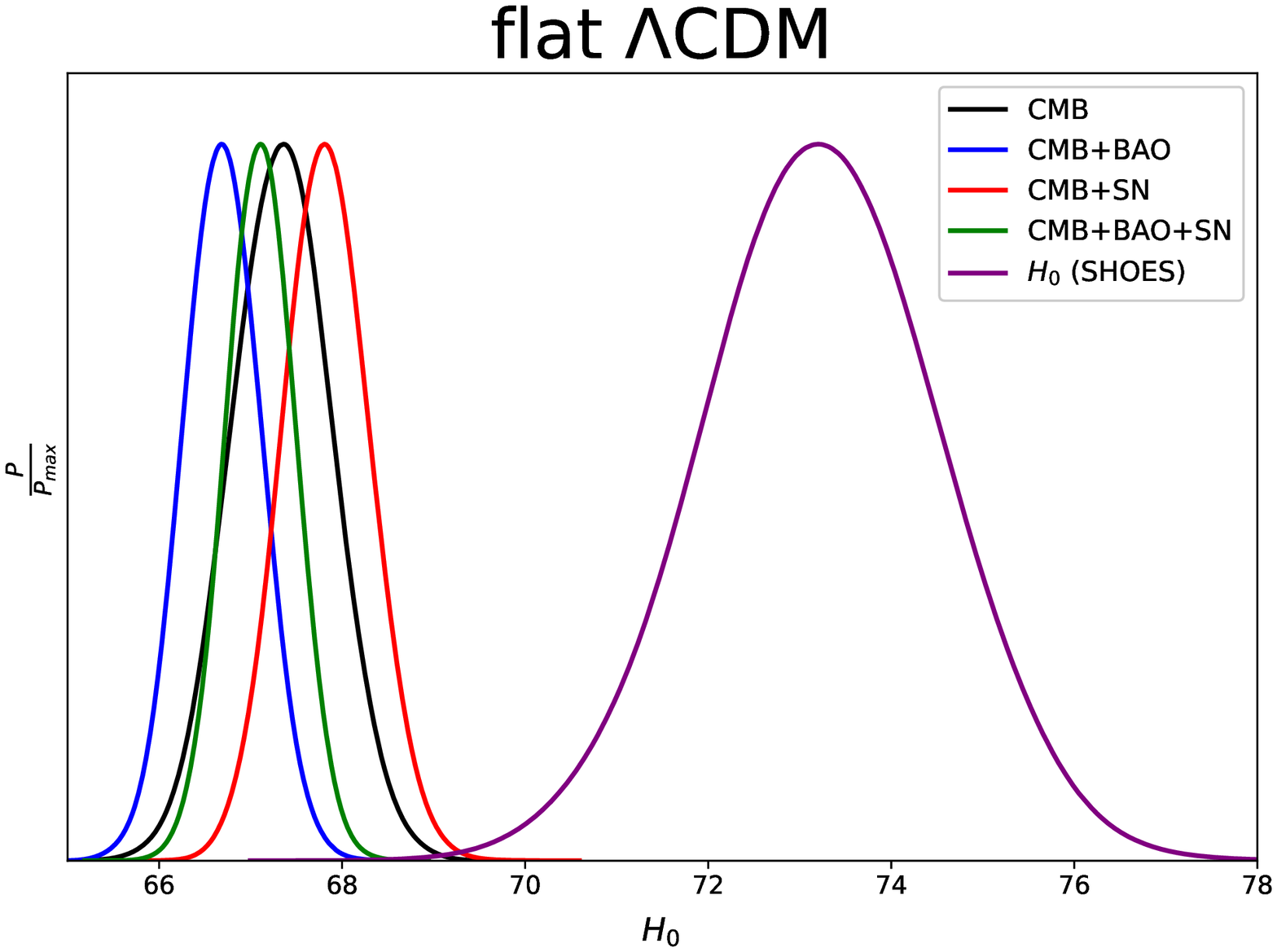}}
\resizebox{250pt}{140pt}{\includegraphics{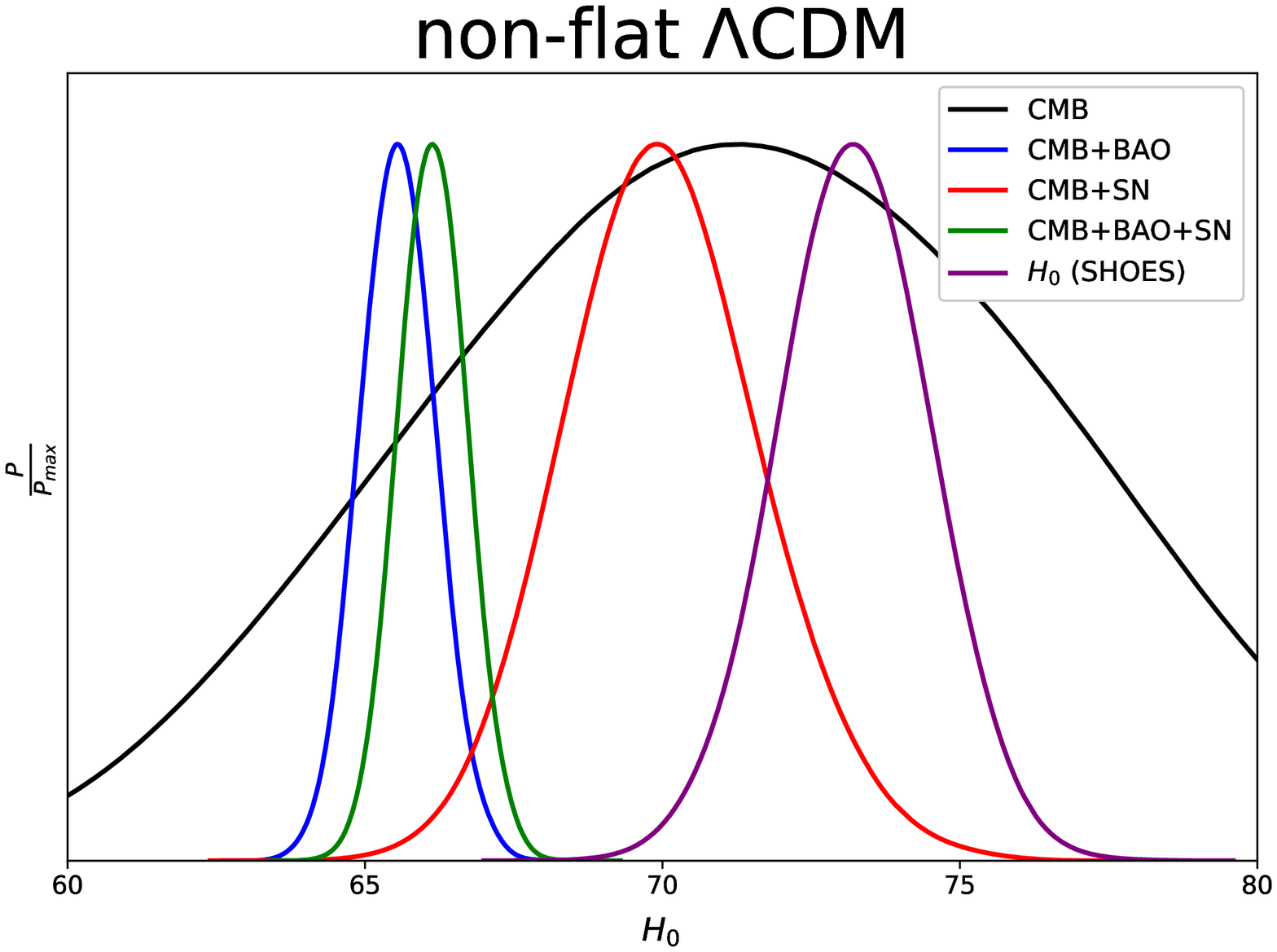}}\\
\resizebox{250pt}{140pt}{\includegraphics{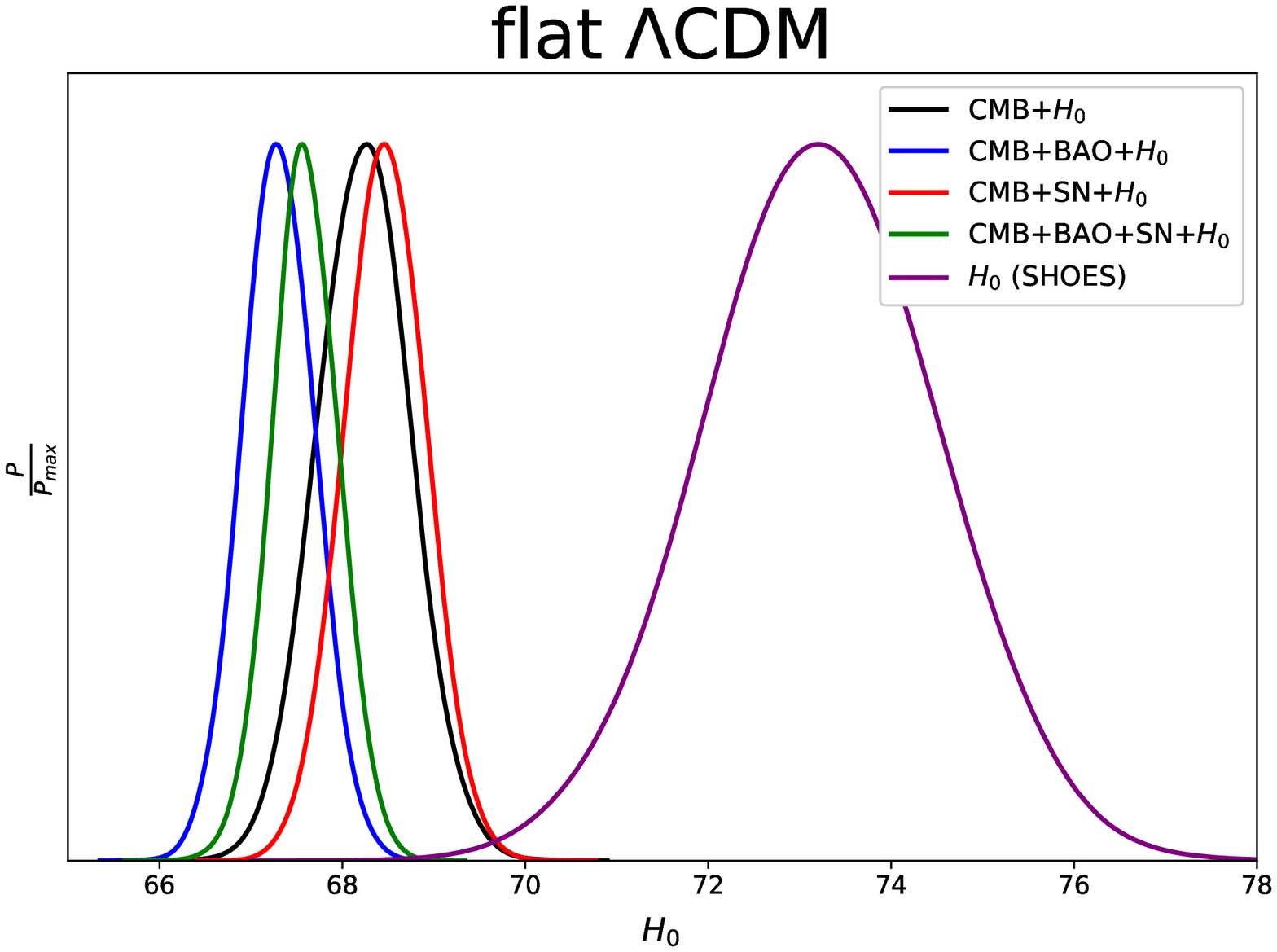}}
\resizebox{250pt}{140pt}{\includegraphics{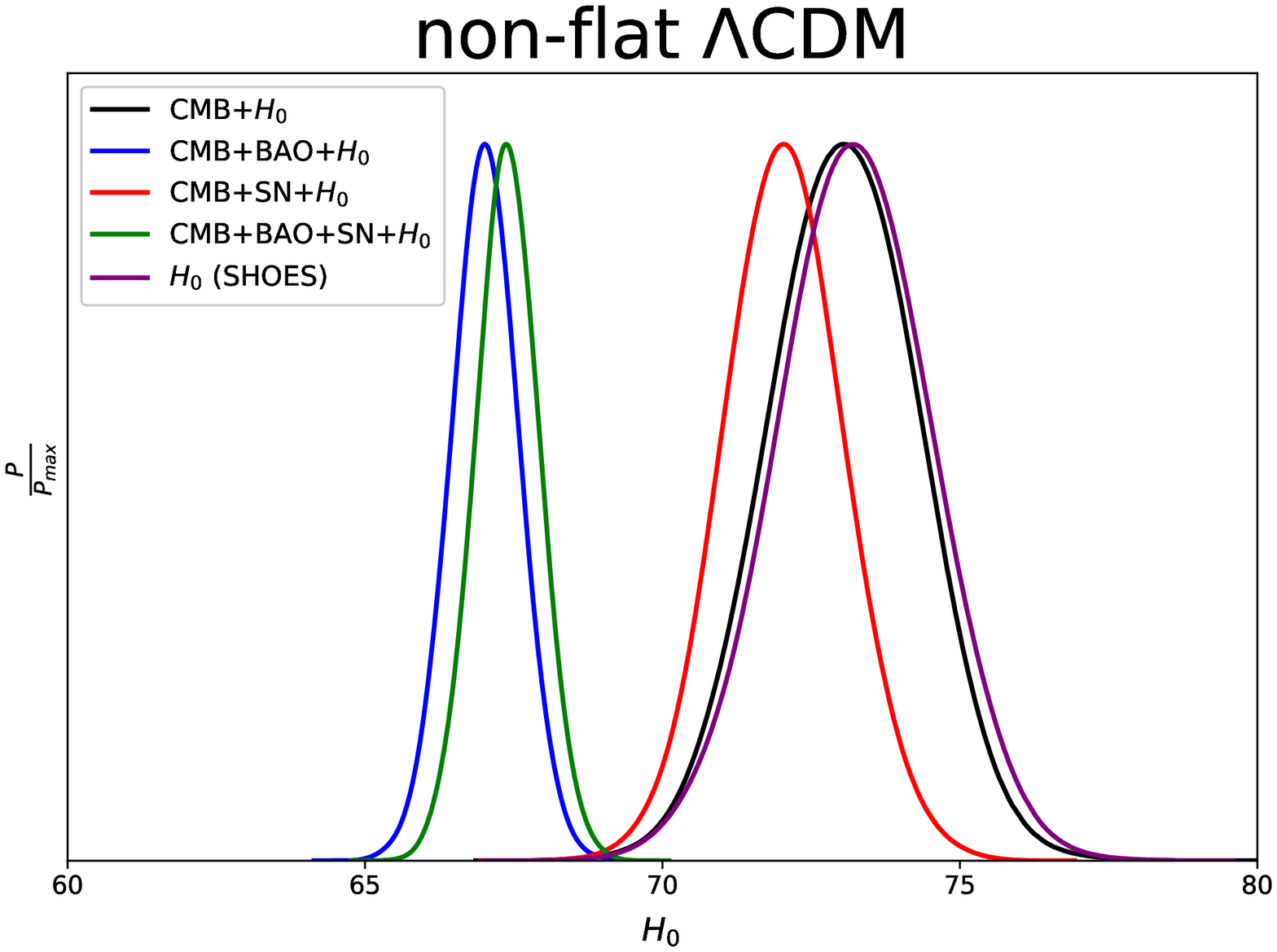}}
\caption{Hubble tension between CMB data (and with other combinations of data sets) and SHOES observation. The left and right panels correspond to the base flat and non-flat $\Lambda$CDM models respectively.}
\label{fig:H0Lcdm}
\end{center}
\end{figure*}

\section{Can late time modification solve the Hubble tension?}
\label{sec-H0tension}

In this section, we discuss the Hubble tension (i.e. the so-called $H_{0}$ discrepancies between early time observations like CMB \citep{Aghanim:2018eyx} and late time local distance ladder observations like SHOES \citep{Riess:2020fzl}) in detail.

In Fig.~\ref{fig:H0Lcdm}, we show the Hubble tension i.e. discrepancies in $H_{0}$ values between CMB data (and with other combinations of data sets) and its local distance ladder measurement from SHOES in the base $\Lambda$CDM model. The left and right panels correspond to the flat and non-flat $\Lambda$CDM models respectively. The $H_{0}$ value and its standard deviation corresponding to the SHOES observation is given by \citep{Riess:2020fzl}

\begin{equation}
H_{0} = 73.2 \pm 1.3.
\label{eq:SHOES}
\end{equation}

In Fig.~\ref{fig:H0Lcdm}, the x-axis shows the $H_{0}$ value and the y-axis shows its normalized probability distribution, obtained from different combinations of data sets. Black, blue, red, green, and purple lines correspond to CMB, CMB+BAO, CMB+SN, CMB+BAO+SN, and SHOES data respectively. The upper panels correspond to the data analysis without any $H_{0}$ prior. The lower panels correspond to the addition of $H_{0}$ prior from SHOES observation, mentioned in Eq.~\eqref{eq:SHOES}. Note that we use Eq.~\eqref{eq:SHOES} as a Gaussian prior for $H_{0}$. From the upper-left panel, we can see that, for the flat $\Lambda$CDM model, the central values of $H_{0}$ corresponding to CMB, CMB+BAO, CMB+SN, CMB+BAO+SN data are around 4.5$\sigma$, 5.0$\sigma$, 4.1$\sigma$ and 4.7$\sigma$ away from the corresponding central value of SHOES observation respectively. Comparing the upper-left and the lower-left plots, we can see that, even with the addition of the $H_{0}$ prior, the tension does not improve significantly.

In Fig.~\ref{fig:H0Lcdm}, in the right panels, we vary the spatial curvature, $\Omega_{\text{k0}}$. Interestingly, from the upper-right panel, we can see that the 1$\sigma$ region corresponding to CMB data overlaps with the 1$\sigma$ region corresponding to the SHOES observation. For the CMB+SN combination of data sets, the tension decreases a bit from 4.1$\sigma$ to 2.5$\sigma$, when we vary the $\Omega_{\text{k0}}$ parameter. For the other two combinations of data sets, the tension does not improve significantly. That means the inclusion of BAO data constrain the $H_{0}$ parameter more significantly compared to CMB and SN data. This fact can be seen more clearly in the lower-right panel when we add the $H_{0}$ prior. We can see that the $H_{0}$ tension vanishes for the CMB alone data when we vary $\Omega_{\text{k0}}$ (without increasing any dark energy degrees of freedom). The $H_{0}$ tension reduces significantly to 0.9$\sigma$ for CMB+SN data. But the tension does not decrease significantly when we include the BAO data.

Next, we want to see if late time modification of the cosmic expansion (i.e. by increasing the dark energy degrees of freedom) can solve the Hubble tension or not. To do this, we use our parametrization for different $d$ values to check how much tension can be decreased for each data combination given by CMB+$H_{0}$, CMB+BAO+$H_{0}$, CMB+SN+$H_{0}$, and CMB+BAO+SN+$H_{0}$. And for any of these data combinations, if Hubble tension vanishes, we check for which value of $d$ it happens so. To do so, we study both the flat and non-flat cases separately to check the importance of the spatial curvature term.

\begin{figure}[]
\begin{center}
\resizebox{250pt}{140pt}{\includegraphics{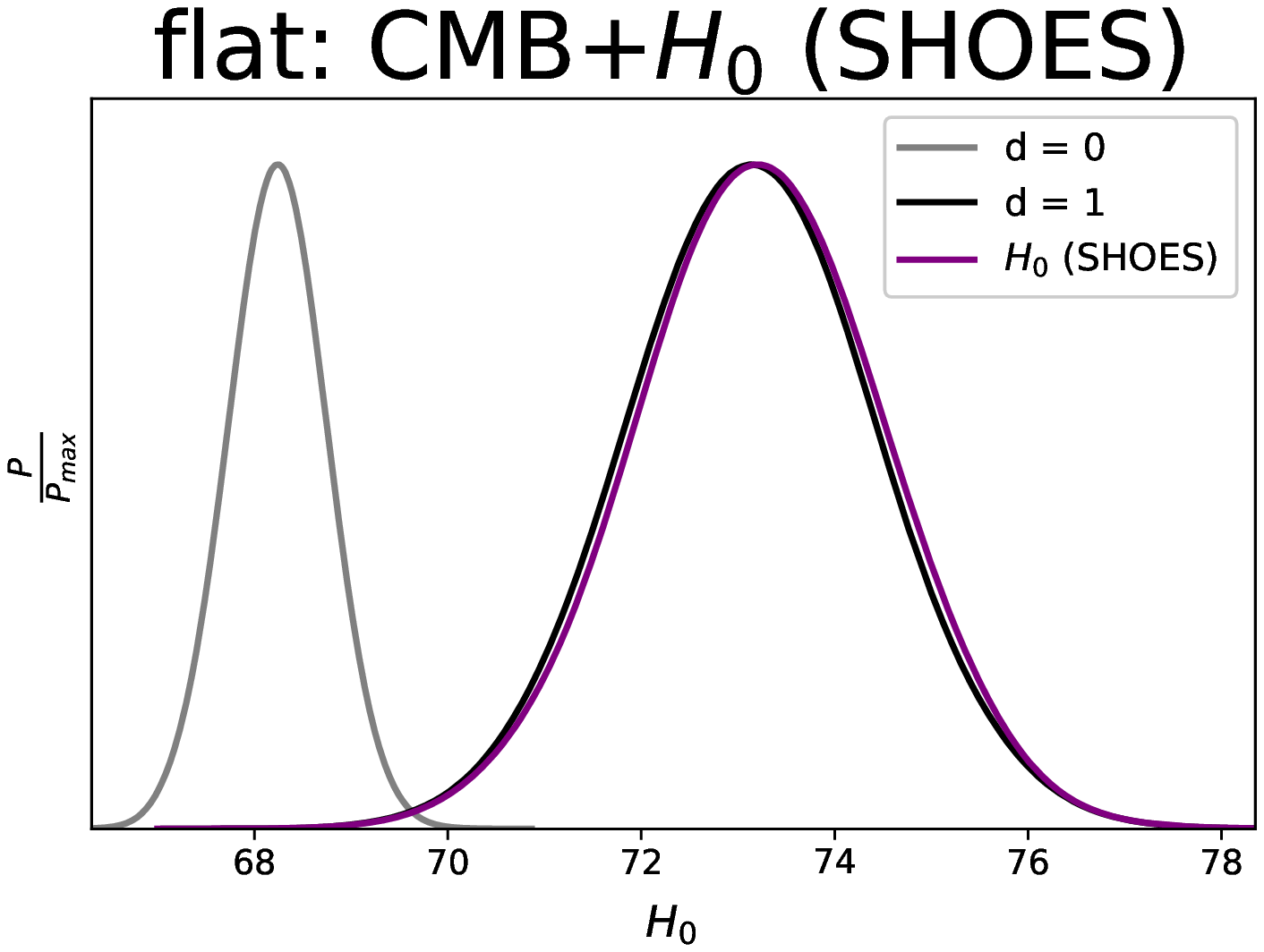}}\\
\resizebox{250pt}{140pt}{\includegraphics{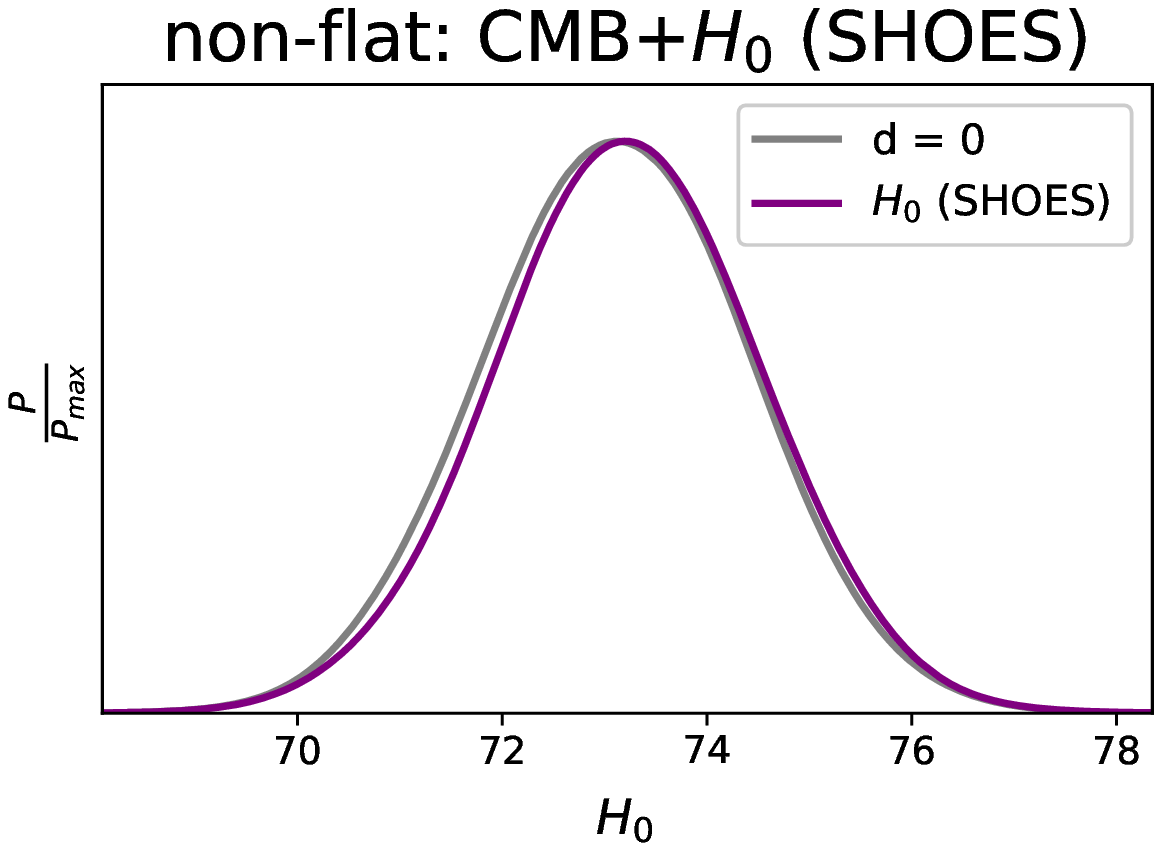}}
\caption{Normalized probability distribution of $H_{0}$ from CMB+$H_{0}$ data with different $d$ values. Purple line corresponds to the SHOES measurement of $H_{0}$.}
\label{fig:CMB_H0}
\end{center}
\end{figure}

\begin{figure}[]
\begin{center}
\resizebox{250pt}{140pt}{\includegraphics{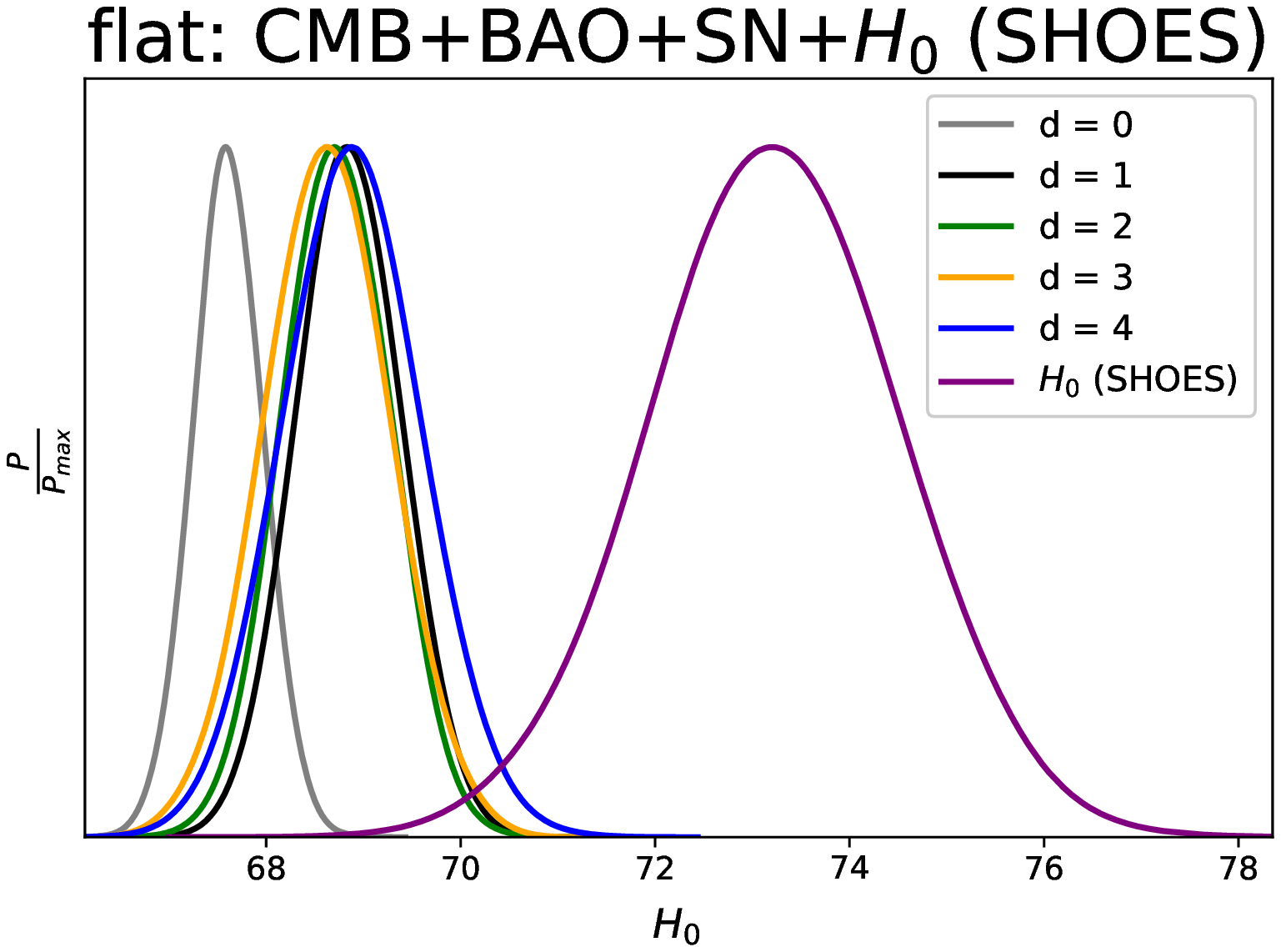}}\\
\resizebox{250pt}{140pt}{\includegraphics{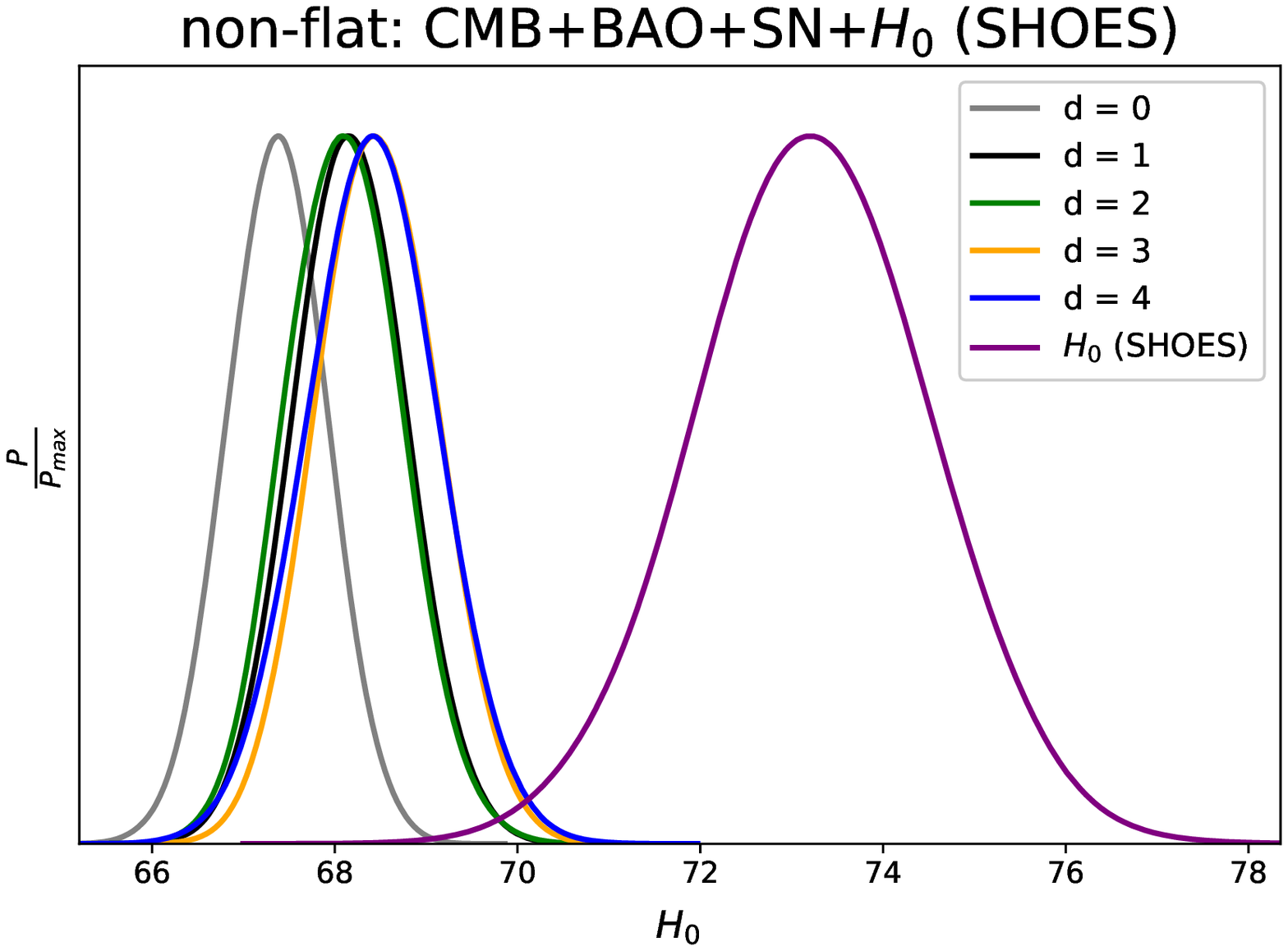}}
\caption{Normalized probability distribution of $H_{0}$ from CMB+BAO+SN+$H_{0}$ data with different $n$ values. Purple line corresponds to the SHOES measurement of $H_{0}$.}
\label{fig:CMB_BAO_SN_H0}
\end{center}
\end{figure}

In Fig.~\ref{fig:CMB_H0}, we plot the normalized probability distribution of $H_{0}$ for the CMB+$H_{0}$ data both for flat and non-flat cases. In this figure, gray, black, green, orange, and blue colors represent $d=0$, 1, 2, 3, and 4 respectively and the purple color corresponds to the SHOES value of $H_{0}$. We also follow the same color code till Fig.~\ref{fig:CMB_BAO_SN_H0}. The upper panel shows that for the flat Universe, Hubble tension can be solved for $d=1$ (i.e with dark energy degree of freedom $1$) or above. From the lower panel, we can see that the Hubble tension can be solved with $d=0$ (corresponding to no dark energy degrees of freedom) when we vary $\Omega_{\text{k0}}$. This is the case we have seen in the previous figure (from the lower-right panel in Fig.~\ref{fig:H0Lcdm}) for the non-flat $\Lambda$CDM, which is a model with no dark energy degrees of freedom.

\begin{figure}[]
\begin{center}
\resizebox{250pt}{140pt}{\includegraphics{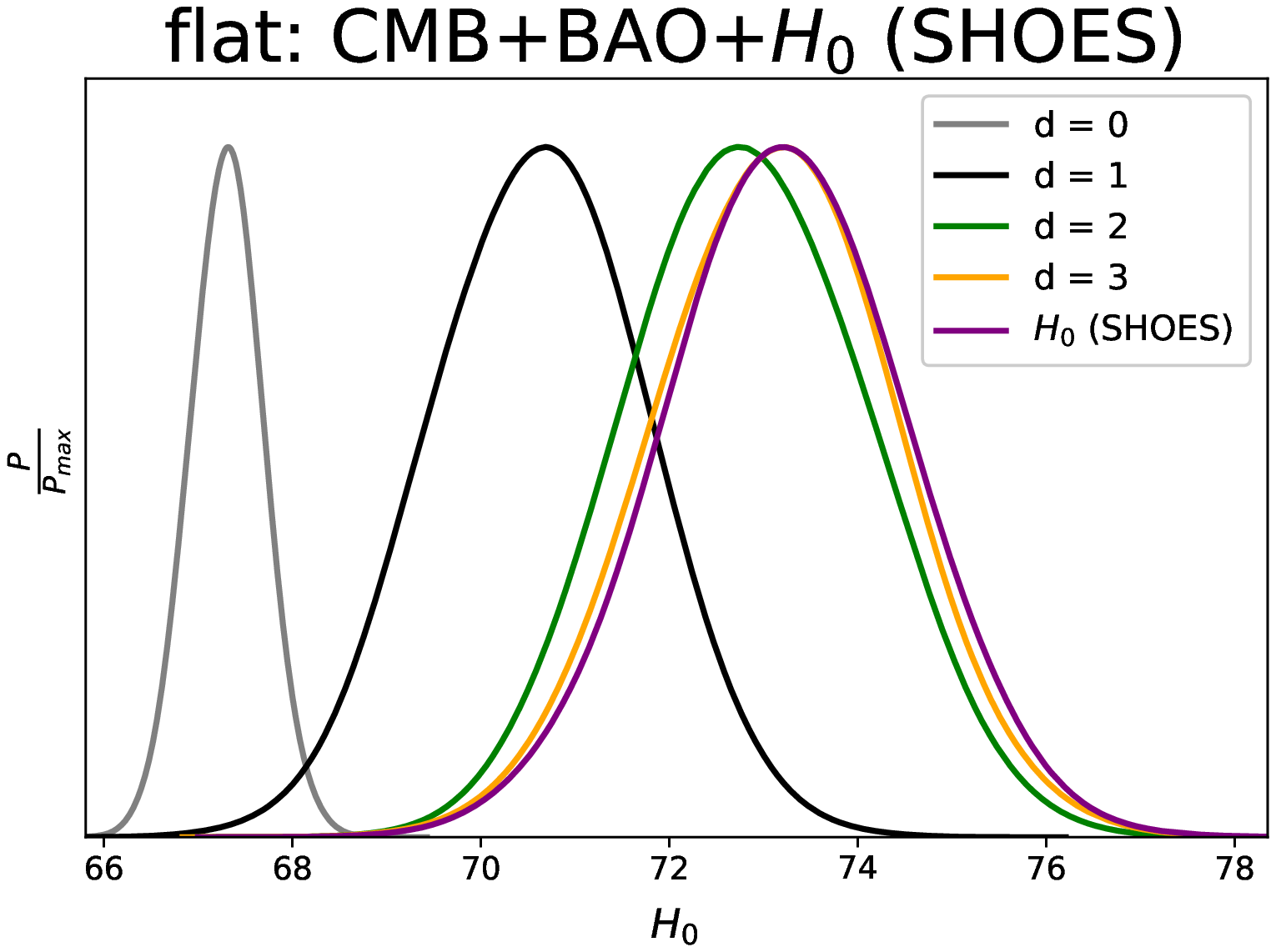}}\\
\resizebox{250pt}{140pt}{\includegraphics{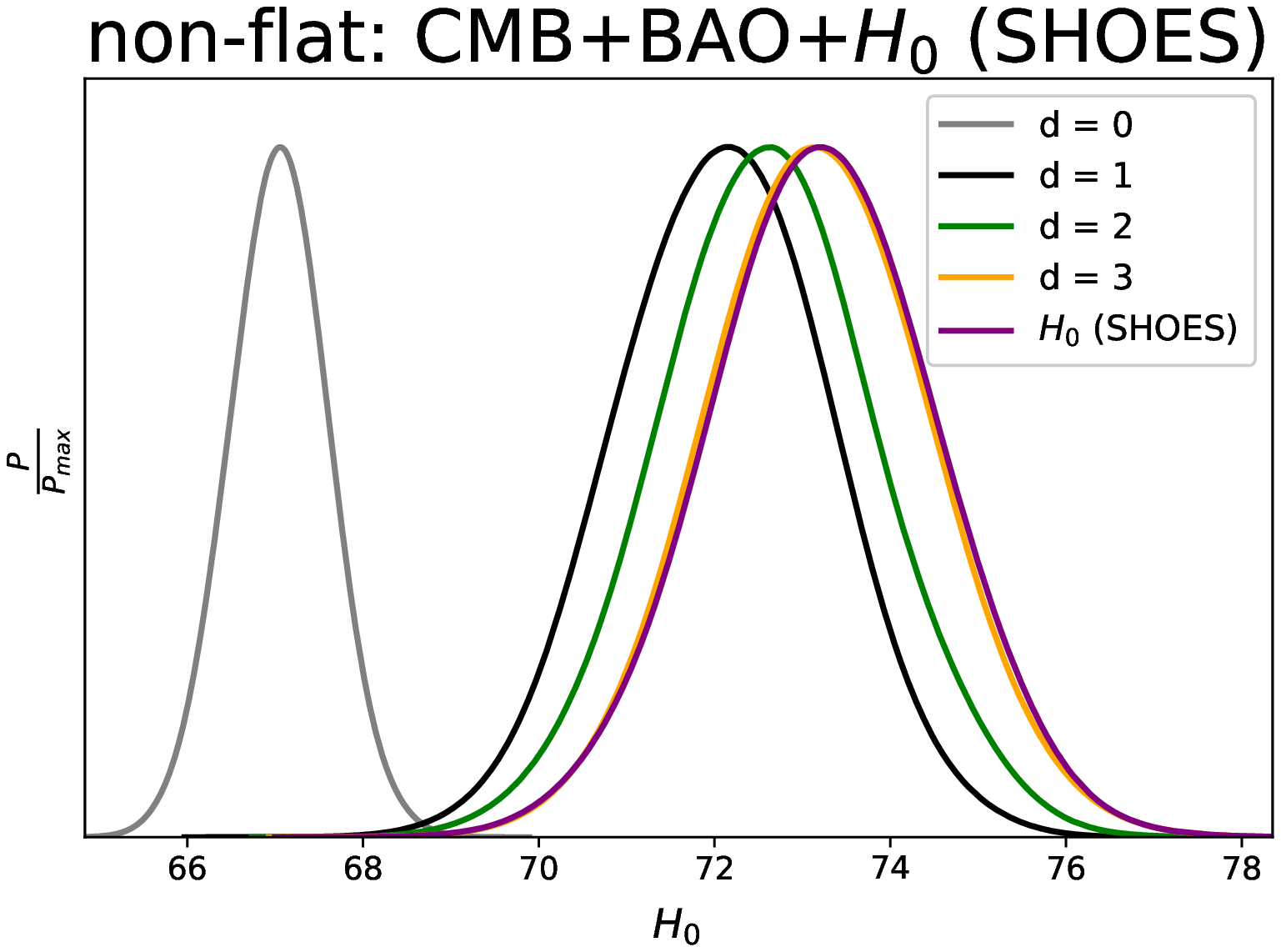}}
\caption{Normalized probability distribution of $H_{0}$ from CMB+BAO+$H_{0}$ data with different $d$ values. Purple line corresponds to the SHOES measurement of $H_{0}$.}
\label{fig:CMB_BAO_H0}
\end{center}
\end{figure}

In Fig.~\ref{fig:CMB_BAO_H0}, we show the normalized probability distribution of $H_{0}$ obtained from CMB+BAO+$H_{0}$ data and compare it with the corresponding SHOES observation. Upper and lower panels correspond to flat and non-flat cases. Interestingly, with the inclusion of BAO data the probability distributions of $H_{0}$ are not significantly different in flat and non-flat cases. That means the BAO data put a strong constraint on the $\Omega_{\text{k0}}$ parameter. From this figure, we find that in this case, the Hubble tension can be solved for $d=3$ (i.e. dark energy degrees of freedom $3$) or above.

\begin{figure}[]
\begin{center}
\resizebox{250pt}{140pt}{\includegraphics{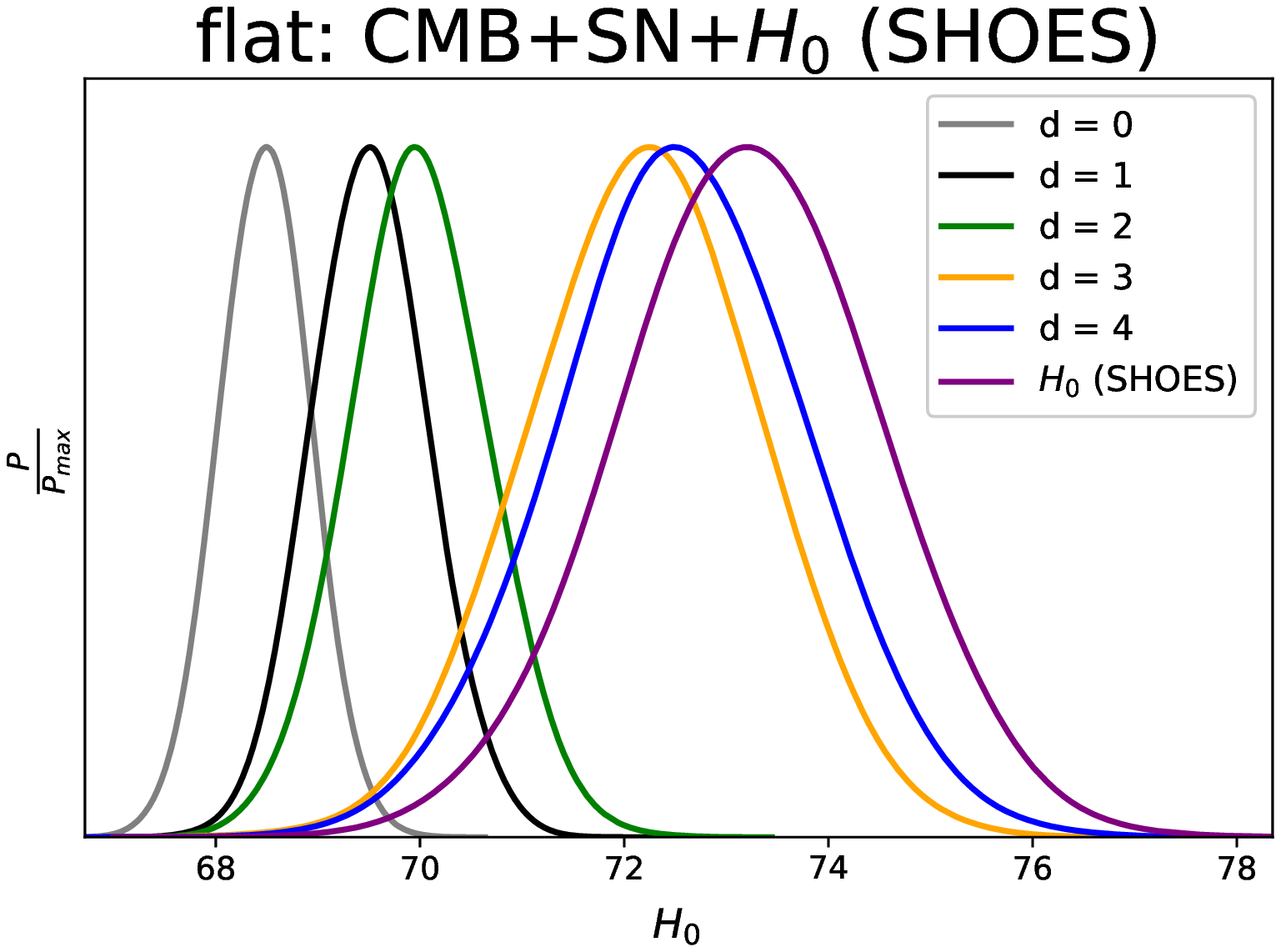}}\\
\resizebox{250pt}{140pt}{\includegraphics{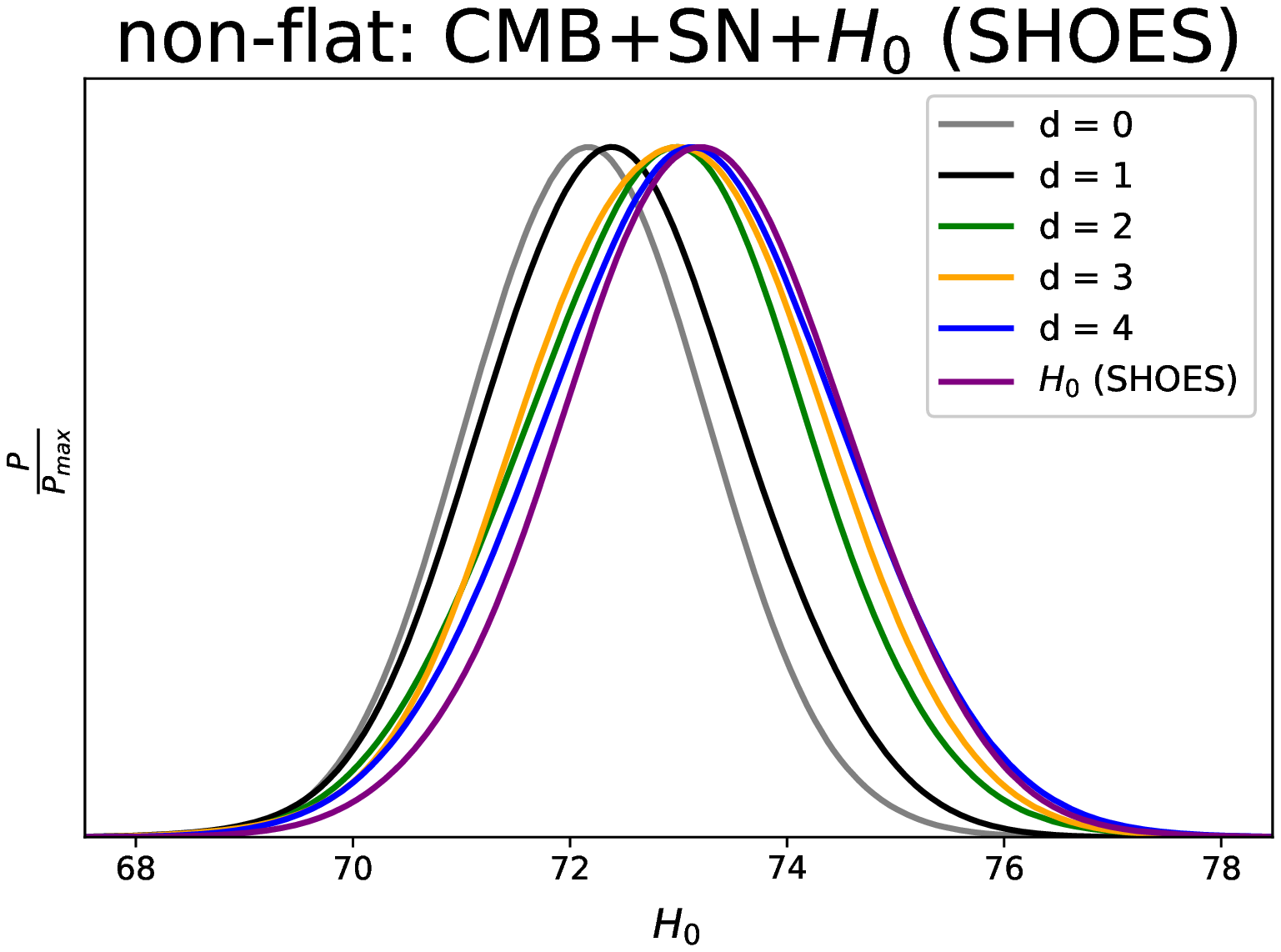}}
\caption{Normalized probability distribution of $H_{0}$ from CMB+SN+$H_{0}$ data with different $d$ values. Purple line corresponds to the SHOES measurement of $H_{0}$.}
\label{fig:CMB_SN_H0}
\end{center}
\end{figure}

In Fig.~\ref{fig:CMB_SN_H0}, we show the normalized probability distribution of $H_{0}$ obtained from CMB+SN+$H_{0}$ data and compare it with the corresponding SHOES observation. For the flat case (from the upper panel), the Hubble tension does not completely vanish but it significantly decreases with increasing $d$. On the other hand, for the non-flat case (from the lower panel), the Hubble tension vanishes for around $d=4$ (i.e. dark energy degrees of freedom $4$) or above.

In Fig.~\ref{fig:CMB_BAO_SN_H0}, we show the normalized probability distribution of $H_{0}$ obtained from CMB+BAO+SN+$H_{0}$ data and compare it with the corresponding SHOES observation. Both for flat and non-flat cases, we can see that the Hubble tension does not significantly decrease for any value of $d$ for the CMB+BAO+SN+$H_{0}$ data. Note that we have checked this fact only up to $d=4$ i.e. up to dark energy degrees of freedom $4$. Although we have studied the Hubble tension for $d$ values up to $4$, from the trend in Fig.~\ref{fig:CMB_BAO_SN_H0}, it is clear that Hubble tension can not be solved even with immediate higher values of $d$ (i.e. higher number of degrees of freedom) except for unexpectedly very large values of $d$. This is almost impossible to do the data analysis with such a large number of parameters and we should not consider a model which possesses such a large number of parameters. So, we can safely say that, for a reasonable good model (i.e a model which does not have a large number of parameters or does not possess any unexpected abrupt changes in behavior at very low redshifts), the Hubble tension can not be solved by the late time modification for the CMB+BAO+SN data.

The conclusion is that the late time modification of the cosmic expansion can solve the Hubble tension for CMB, CMB+BAO, and CMB+SN combinations of data sets, but interestingly, when three data are combined i.e. for CMB+BAO+SN combination of data sets, the late time modification can not solve the Hubble tension.

\section{Is $M$ tension more fundamental than $H_{0}$ tension?}
\label{sec-Mtension}

Recently, some authors like in \citep{Camarena:2021jlr,Efstathiou:2021ocp} have claimed that one should not use the $H_{0}$ prior from distance ladder observations like SHOES, mentioned in Eq.~\eqref{eq:SHOES}. This is because the SHOES observation uses low redshift ($0.0233<z<0.15$) data of type Ia supernova relative magnitude to derive $H_{0}$ from the absolute magnitude $M$. See \citep{Riess:2016jrr,Camarena:2021jlr} for details. So, if we use type Ia supernovae data (Pantheon sample here) and $H_{0}$ prior simultaneously, we use the low redshift supernovae data twice. In this sense, it is wrong. Plus, derivation of $H_{0}$ from $M$ is a model-dependent procedure (although model dependency is not very significant because the redshift range is smaller as $0.0233<z<0.15$). These are the two reasons that one should avoid $H_{0}$ prior and use $M$ prior instead and check whether there are any discrepancies in values of M between early time observation and late time local distance ladder observation. So, $M$ tension should be considered more fundamental than $H_{0}$ tension and we should check if late time modification can solve the $M$ tension or not. See \citep{Camarena:2021jlr} for more detailed discussions about this issue. So, in this section, we use $M$ prior and see how the results change.

The equivalent $M$ prior corresponding to the SHOES observation for the Pantheon sample (in the redshift range $0.0233<z<0.15$) is given by \citep{Camarena:2021jlr}

\begin{equation}
M = -19.2435 \pm 0.0373.
\label{eq:SHOESeqvlntM}
\end{equation}

\begin{figure}[]
\begin{center}
\resizebox{250pt}{140pt}{\includegraphics{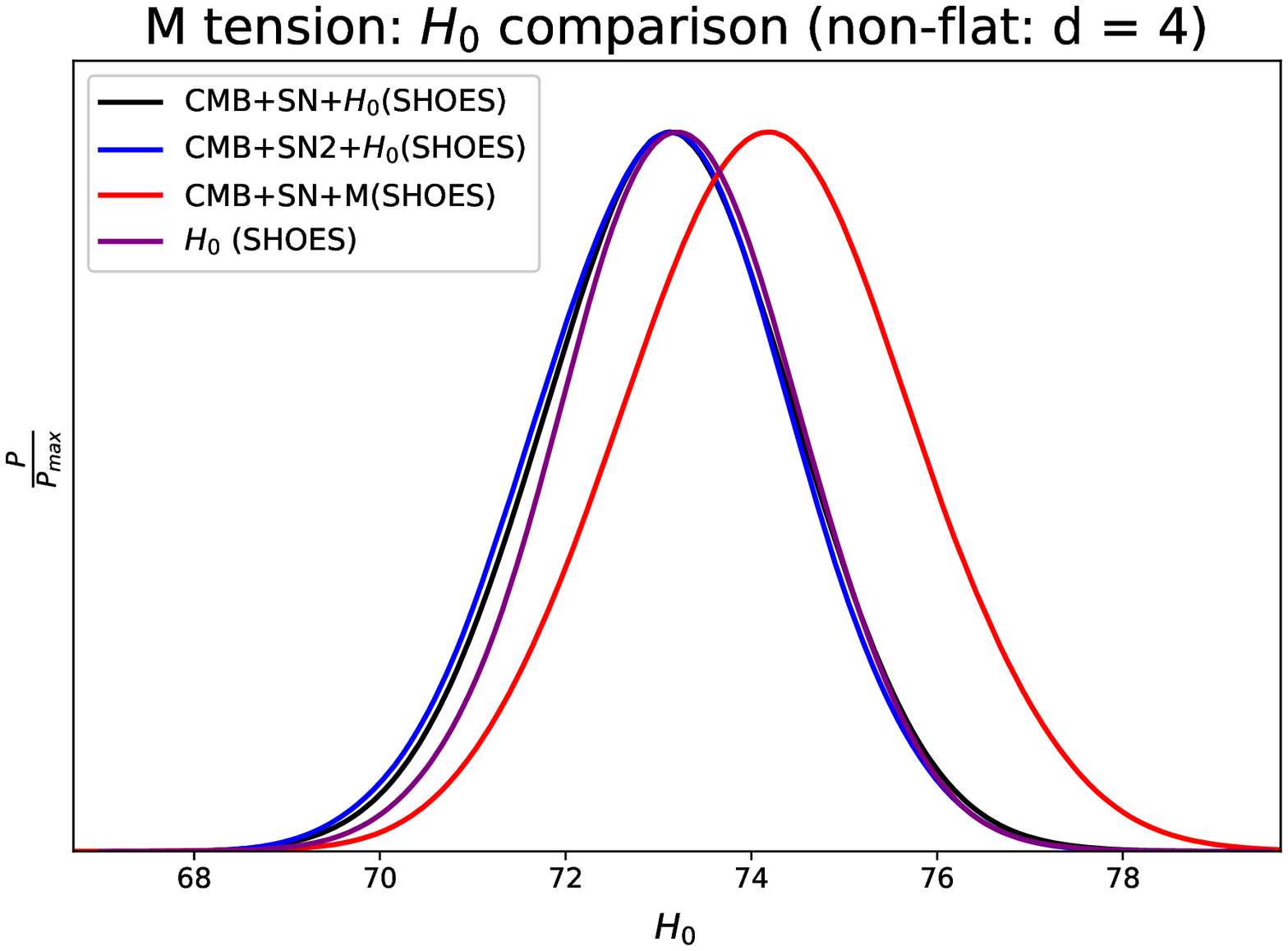}}\\
\resizebox{250pt}{140pt}{\includegraphics{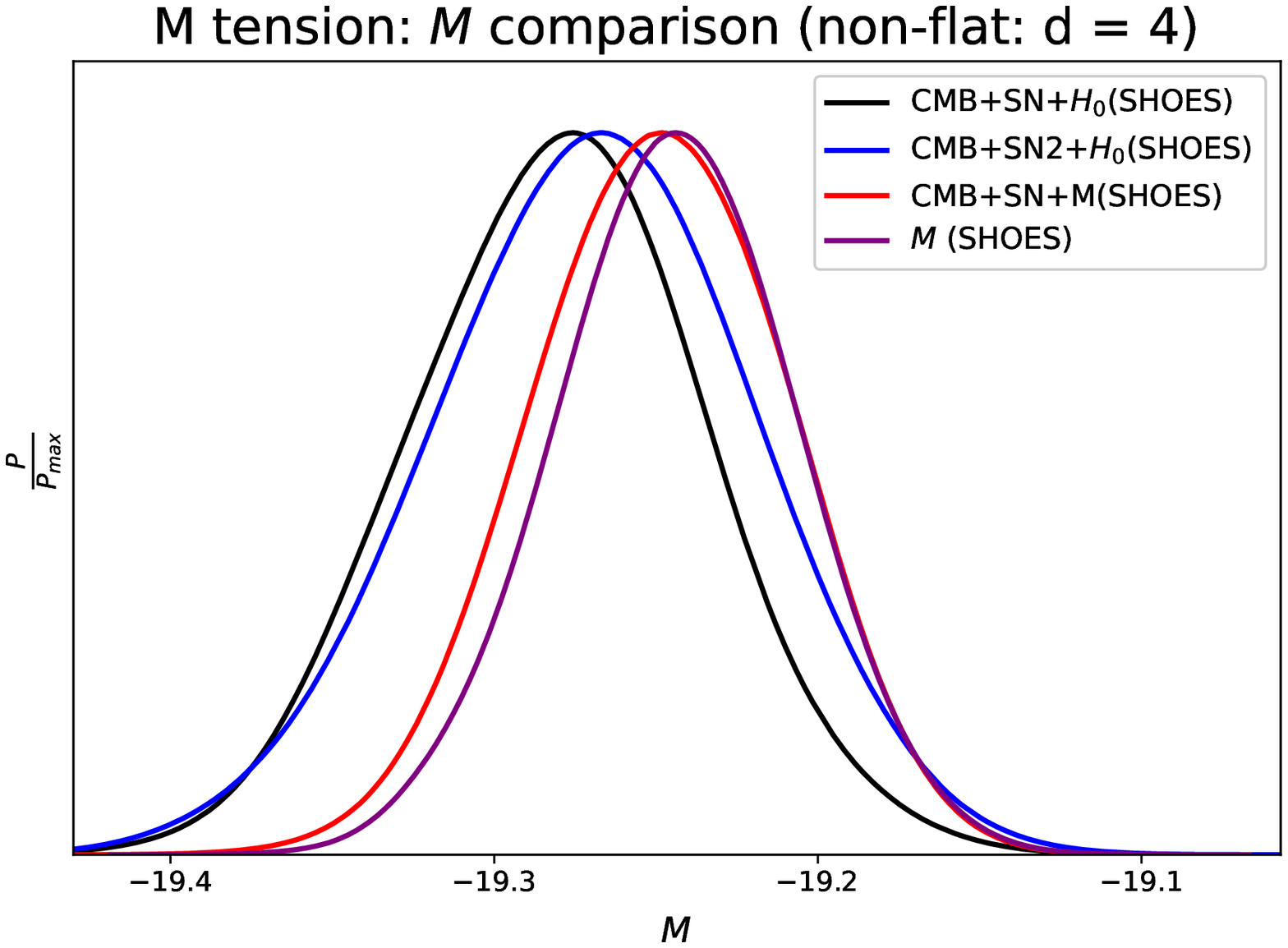}}
\caption{$H_{0}$ and $M$ tension comparison for CMB+SN combination of data sets.}
\label{fig:H0_M_CMB_SN}
\end{center}
\end{figure}

\begin{figure}[]
\begin{center}
\resizebox{250pt}{140pt}{\includegraphics{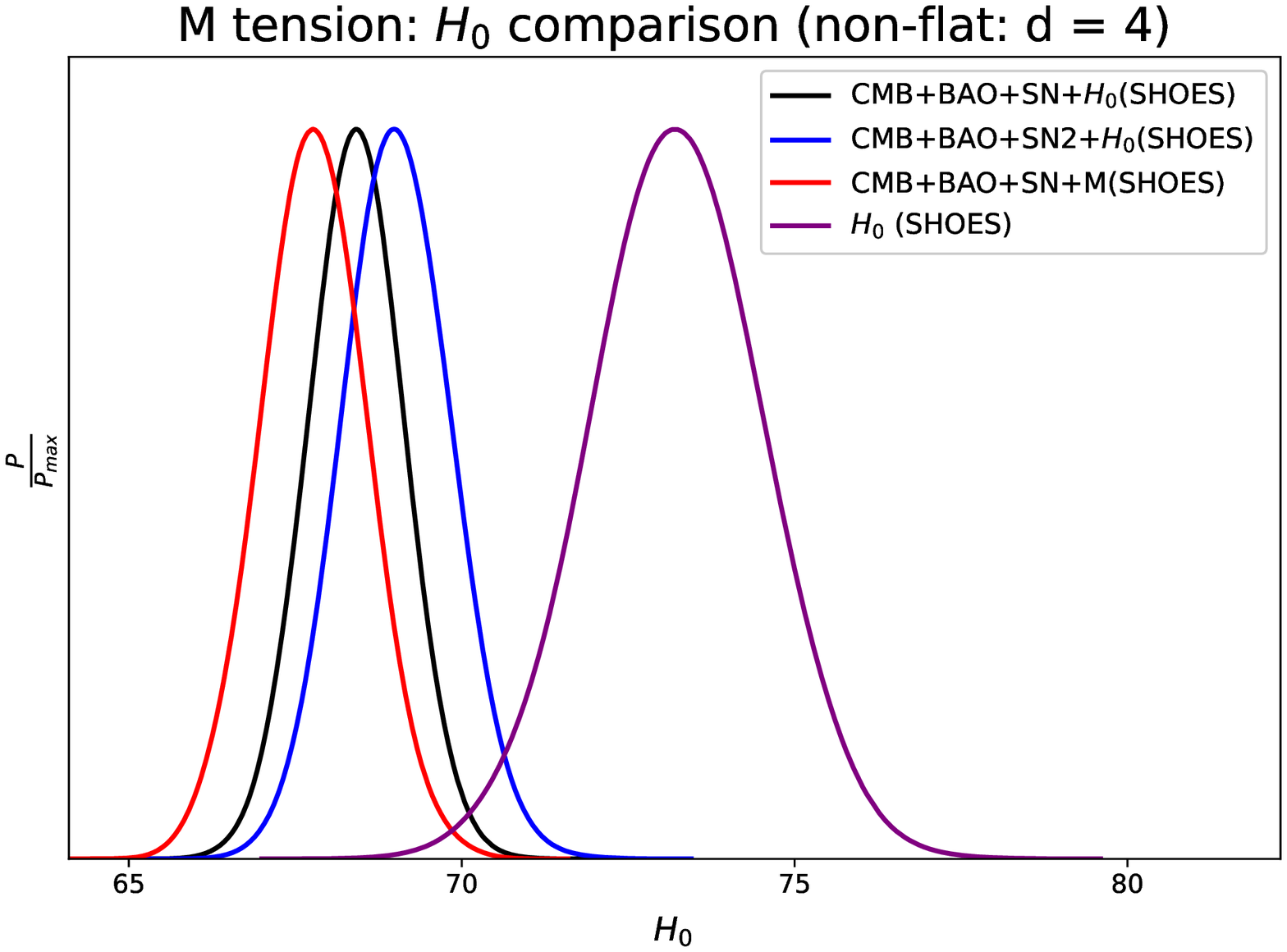}}\\
\resizebox{250pt}{140pt}{\includegraphics{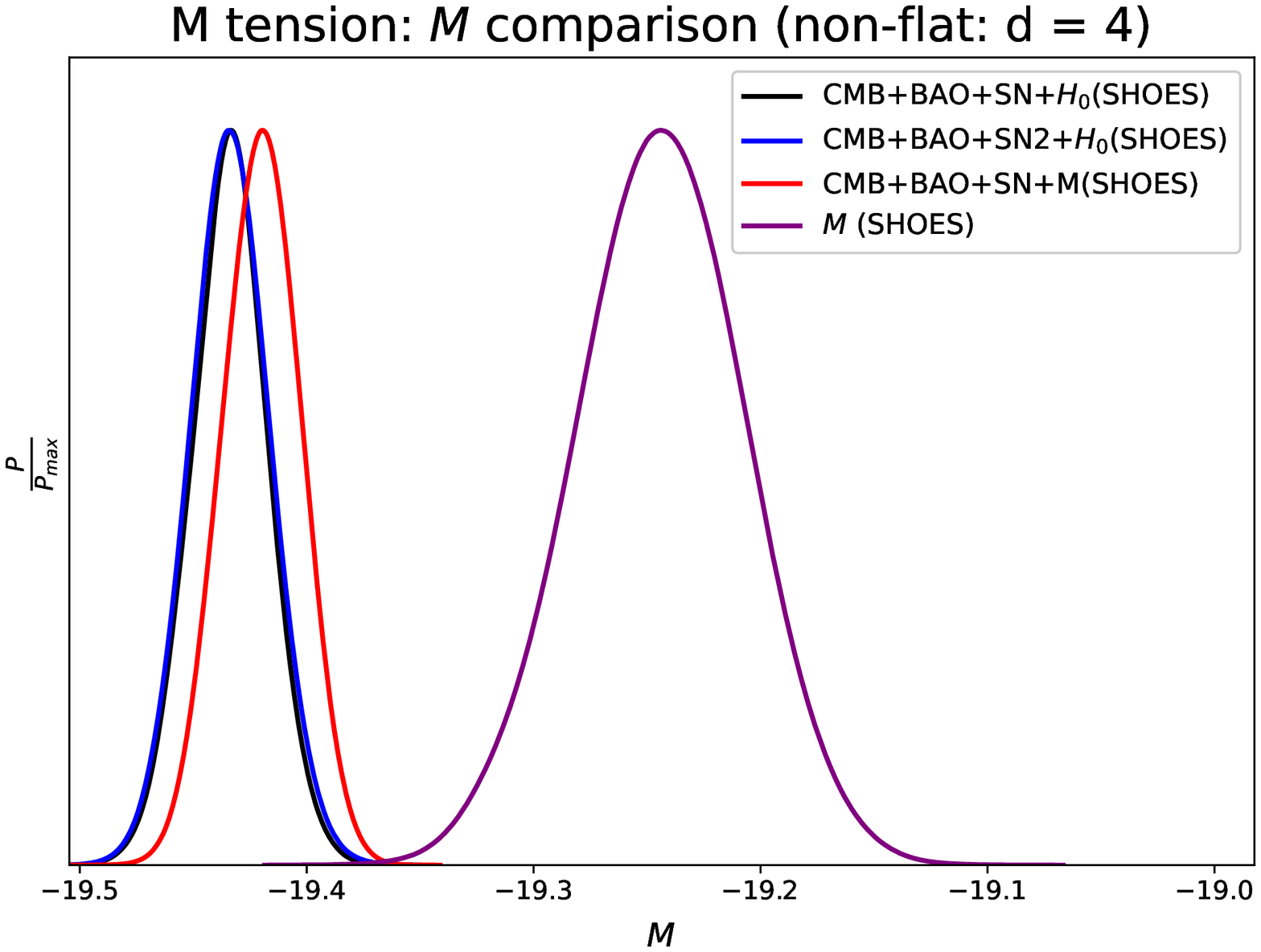}}
\caption{$H_{0}$ and $M$ tension comparison for CMB+BAO+SN combination of data sets.}
\label{fig:H0_M_CMB_BAO_SN}
\end{center}
\end{figure}

For the CMB and CMB+BAO combinations of data sets, the use of $M$ prior is meaningless. It can be used only when Supernova type Ia (here Pantheon Sample) or equivalent data is included in the data analysis. So, we repeat the data analysis for CMB+SN in Fig.~\ref{fig:H0_M_CMB_SN} (corresponding to Fig.~\ref{fig:CMB_SN_H0}) and CMB+BAO+SN in Fig.~\ref{fig:H0_M_CMB_BAO_SN} (corresponding to Fig.~\ref{fig:CMB_BAO_SN_H0}) combinations of data sets respectively with $M$ prior. In Figs.~\ref{fig:H0_M_CMB_SN} and~\ref{fig:H0_M_CMB_BAO_SN}, we compare three cases given below.

\begin{itemize}
\item
(1) In the first case, we keep $H_{0}$ prior as usual i.e. same as in the previous section. We represent this by dashed lines.
\item
(2) In the second case, we keep $H_{0}$ prior but exclude data of the range ($0.0233<z<0.15$) from Pantheon Sample and we denote this sample as 'SN2'. We represent this by dotted lines.
\item
(3) Finally, in the third case, we use $M$ prior instead of $H_{0}$. We denote this by solid lines.
\end{itemize}

In Fig.~\ref{fig:CMB_SN_H0}, we have seen that the non-flat $d=4$ (or above) parametrization can solve the Hubble tension in the case of CMB+SN data. So, we take this model in Fig.~\ref{fig:H0_M_CMB_SN} and compare the three cases mentioned above. In this figure and the next figure, in the top panels, the x-axis and the y-axis correspond to the $H_{0}$ and its normalized probability distribution respectively. Similarly, in the bottom panels, the x-axis and the y-axis correspond to the $M$ and its normalized probability distribution respectively. From the top panel in Fig.~\ref{fig:H0_M_CMB_SN}, we can see that $H_{0}$ follows the SHOES value both for first (CMB+SN+$H_{0}$) and second cases (CMB+SN2+$H_{0}$). It is obvious because we have already seen that (from Fig.~\ref{fig:CMB_SN_H0}), for the case of CMB+SN+$H_{0}$, the $H_{0}$ value already marges with the SHOES $H_{0}$ value. And since SN2 data has a lesser number of data points compared to SN data, it has the lesser constraining power on $H_{0}$, thus, for the case of CMB+SN2+$H_{0}$ data, the $H_{0}$ value would more easily marge with SHOES value of $H_{0}$. The $H_{0}$ value increases a bit for the third case i.e. for CMB+SN+M data. One can check that this fact is consistent with \citep{Camarena:2021jlr}. On the other hand, from the bottom panel, we can see that the $M$ value is lesser in the first case (CMB+SN+$H_{0}$) compared to the third case (CMB+SN+M). The $M$ value for second case (CMB+SN2+$H_{0}$) is further lesser. The third case CMB+SN+M in the bottom panel shows that $M$ tension can also be solved by the non-flat $d=4$ (or above) parametrization. In summary, the results from the three cases are a little bit different but these differences are not very significant.

In Fig.~\ref{fig:H0_M_CMB_BAO_SN}, we repeat the same plot as in Fig.~\ref{fig:H0_M_CMB_SN} but for the CMB+BAO+SN data. Here, we do it for the non-flat $d=4$ parametrization (since it has the highest number of degrees of freedom or the parameters in our analysis). Here, we can see that neither $H_{0}$ tension nor the $M$ tension can be solved for $d\leq 4$ parametrization. Other behaviors are similar as in the previous figure. One extra fact to learn from this figure is that when CMB, BAO, and SN data are combined, the $M$ tension is slightly larger compared to the $H_{0}$ tension. To mention, the $H_{0}$ tensions are at 3.7$\sigma$, 3.2$\sigma$, and 4.2$\sigma$ confidence levels for CMB+BAO+SN+$H_{0}$, CMB+BAO+SN2+$H_{0}$, and CMB+BAO+SN+M respectively, whereas the $M$ tensions are at 5.1$\sigma$, 5.1$\sigma$, and 4.7$\sigma$ confidence levels for CMB+BAO+SN+$H_{0}$, CMB+BAO+SN2+$H_{0}$, and CMB+BAO+SN+M respectively. So, the $M$ tension is a little bit larger than the Hubble tension.

\section{Conclusion}
\label{sec-conclusion}

We present an analytical parametrization to the line of sight comoving distance and the normalized Hubble parameter to study the late time modification of the cosmic expansion beyond the $\Lambda$CDM model. This parametrization includes the contribution from spatial curvature terms as well as it captures accurate higher redshift behaviors as well. In this way, all the background quantities (related to CMB, BAO, and SN data) become analytic. Thus, it is easier to implement in the data analysis.

With this parameterization, we put constraints on important background cosmological parameters like $h$, $\Omega_{\text{m0}}$, $\Omega_{\text{k0}}$, $\Omega_{\text{b0}}$, and $M$. Constraint on $\Omega_{\text{b0}}$ comes when CMB and BAO data are included and the constraint on $M$ comes when SN data is included. We find that CMB, BAO, and SN data combined put significant constraints on the background evolution of the Universe.

We also check if late time modification of the cosmic expansion can solve the so-called Hubble tension between early Universe observation and the late time local distance ladder observations like SHOES. We find that this tension can be solved between CMB $\&$ SHOES, between CMB+BAO $\&$ SHOES, and between CMB+SN $\&$ SHOES by the late time modification. But, when CMB, BAO, and SN data are combined and compared with SHOES, the Hubble tension can not be solved by late time modification. This is because the CMB+BAO+SN combination of data sets put strong enough constraints on background evolution and hence $H_{0}$, so the introduction of SHOES $H_{0}$ prior does not significantly pull the $H_{0}$ value towards the corresponding SHOES value.

Recently, some authors like in \citep{Camarena:2021jlr,Efstathiou:2021ocp} have claimed that one should consider $M$ prior instead of $H_{0}$ prior and try to solve the $M$ tension instead of $H_{0}$ tension between early and local cosmological observations. This is because local distance ladder observations like SHOES already use low redshift ($0.0233<z<0.15$) type Ia supernova data to derive $H_{0}$ from $M$ with some parametrized models of the luminosity distance, so when we use SN data and $H_{0}$ prior simultaneously, we use the low redshift SN data twice. Also, the derivation of $H_{0}$ is not model-independent. Although the model dependency on such low redshifts is not that significant, for accurate results, we can not ignore it.

So, with this motivation, we also replace the $H_{0}$ prior by $M$ prior and check if late time modification can solve the $M$ tension between early and late time local observations. For the case of CMB and CMB+BAO data, it is meaningless to use $M$ prior or try to solve $M$ tension, since parameter $M$ is not involved in these two data. But, it is involved in SN data, so we check the $M$ tension for the two cases, CMB+SN and CMB+BAO+SN with the $M$ prior. We find that the late time modification can solve the $M$ tension between CMB+SN $\&$ SHOES but can not solve it between CMB+BAO+SN $\&$ SHOES. This is because CMB+BAO+SN data combined put tight constraints on $M$ value as well.

For the CMB+BAO+SN combination of data sets, we also find another interesting fact that the $M$ tension is little bit higher than the corresponding $H_{0}$ tension.

In summary, we find that the late time modification of the cosmic expansion does not solve the Hubble tension nor the $M$ tension when we combine CMB, BAO, and SN data.

\section*{Acknowledgements}
BRD would like to acknowledge DAE, Govt. of India for financial support through Postdoctoral Visiting Fellow through TIFR.

\bibliographystyle{apsrev4-1}
\bibliography{refmdlindpgnrl}

\end{document}